\documentclass[twocolumn,trackchanges]{aastex62}

\usepackage{amsmath}
\usepackage{xspace}

\newcommand{\unitspace}{\ensuremath{\mskip\thinmuskip}}
\newcommand{\msun}{\ensuremath{M_{\odot}}\xspace}

\newcommand{\degree}{\ensuremath{^{\circ}}\xspace}

\newcommand{\nue}{\ensuremath{\nu_{e}}\xspace}
\newcommand{\nubar}{\ensuremath{\bar{\nu}_{e}}\xspace}
\newcommand{\nux}{\ensuremath{\nu_{x}}\xspace}

\newcommand{\dint}{\ensuremath{\unitspace\mathrm{d}}}
\newcommand{\vect}[1]{\boldsymbol{#1}}

\newcommand{\new}[1]{#1}

\received{-}
\revised{-}
\accepted{-}
\submitjournal{ApJ}
\shorttitle{Effects of LESA in 3D Supernova Simulations}
\shortauthors{Glas et al.}

\begin{document}

\title{Effects of LESA in Three-Dimensional Supernova Simulations\\
with Multi-Dimensional and Ray-by-Ray-plus Neutrino Transport}

\correspondingauthor{Robert Glas}
\email{rglas@mpa-garching.mpg.de}

\author[0000-0002-7040-9472]{Robert~Glas}
\affiliation{Max-Planck-Institut f\"ur Astrophysik, Karl-Schwarzschild-Stra{\ss}e 1, D-85748 Garching, Germany}
\affiliation{Physik Department, Technische Universit\"at M\"unchen, James-Franck-Stra{\ss}e 1, D-85748 Garching, Germany}

\author[0000-0002-0831-3330]{H.-Thomas~Janka}
\affiliation{Max-Planck-Institut f\"ur Astrophysik, Karl-Schwarzschild-Stra{\ss}e 1, D-85748 Garching, Germany}

\author{Tobias~Melson}
\affiliation{Max-Planck-Institut f\"ur Astrophysik, Karl-Schwarzschild-Stra{\ss}e 1, D-85748 Garching, Germany}

\author{Georg~Stockinger}
\affiliation{Max-Planck-Institut f\"ur Astrophysik, Karl-Schwarzschild-Stra{\ss}e 1, D-85748 Garching, Germany}
\affiliation{Physik Department, Technische Universit\"at M\"unchen, James-Franck-Stra{\ss}e 1, D-85748 Garching, Germany}

\author[0000-0002-3126-9913]{Oliver~Just}
\affiliation{Astrophysical Big Bang Laboratory, RIKEN Cluster for Pioneering Research, 2-1 Hirosawa, Wako, Saitama 351-0198, Japan}
\affiliation{Max-Planck-Institut f\"ur Astrophysik, Karl-Schwarzschild-Stra{\ss}e 1, D-85748 Garching, Germany}

\begin{abstract}

A set of eight self-consistent, time-dependent supernova (SN) simulations in three 
spatial dimensions (3D) for 9\,$M_\odot$ and 20\,$M_\odot$ progenitors is evaluated for 
the presence of dipolar asymmetries of the electron lepton-number emission as discovered
by Tamborra et al.\ and termed lepton-number emission self-sustained asymmetry (LESA).
The simulations were performed with the \textsc{Aenus-Alcar} neutrino/hydrodynamics code,
which treats the energy- and velocity-dependent transport of neutrinos of all flavors 
by a two-moment scheme with algebraic M1 closure. For each of the progenitors, 
results with fully multi-dimensional (FMD) neutrino transport and with
ray-by-ray-plus (RbR+) approximation are considered for two different grid resolutions.
While the 9\,$M_\odot$ models develop explosions, the 20\,$M_\odot$ 
progenitor does not explode with the employed version of simplified neutrino opacities.
In all 3D models we observe the growth of substantial dipole amplitudes of the
lepton-number (electron neutrino minus antineutrino) flux with stable or slowly
time-evolving direction and overall properties fully consistent with the LESA 
phenomenon. Models with RbR+ transport develop LESA dipoles
somewhat faster and with temporarily higher amplitudes, but the FMD calculations 
exhibit cleaner hemispheric asymmetries with a far more dominant dipole. In contrast,
the RbR+ results display much wider multipole spectra of the neutrino-emission 
anisotropies with significant power also in the quadrupole and higher-order modes.
Our results disprove speculations that LESA is a numerical artifact of RbR+ transport.
We also discuss LESA as consequence of a dipolar convection flow inside of the nascent
neutron star and establish, tentatively, a connection to Chandrasekhar's linear theory
of thermal instability in spherical shells.

\end{abstract}

\keywords{convection --- hydrodynamics --- instabilities --- methods: numerical --- neutrinos ---\\ supernovae: general}

\section{Introduction}
\label{sec:introduction}

\citet{2014ApJ...792...96T} and \citet{2016ARNPS..66..341J} reported the 
discovery of a large hemispheric asymmetry of the total electron 
lepton-number emission from the newly formed neutron star in a set of
three-dimensional (3D) core-collapse supernova (CCSN) simulations,
which made use of the ray-by-ray-plus (RbR+) approximation to treat
the multi-dimensionality of the neutrino transport, thereby ignoring
the nonradial components of the neutrino flux vector.
They termed this stunning, new phenomenon 
``lepton-number emission self-sustained asymmetry'' (LESA).
It is characterized by a substantial amplitude of the dipole component 
(in some cases and phases even exceeding the monopole) of the lepton-number flux,
which is defined by the difference of the radial number flux of electron 
neutrinos and antineutrinos at spatial point $\vect{r}$,
\begin{equation}
F_{\mathrm{lnf}} (\vect{r}) =
F^{r,\mathrm{n}}_{\nu_{e}} (\vect{r}) -
F^{r,\mathrm{n}}_{\bar{\nu}_{e}} (\vect{r}).
\end{equation}
The radial number flux of any neutrino species $\nu$ is given by
\begin{equation}\label{eqn:radial_number_flux}
F^{r,\mathrm{n}}_{\nu} =
\int \frac{F^{r}_{\nu} (\varepsilon) }{\varepsilon} \dint{\varepsilon} ,
\end{equation}
with the radial energy-flux density ${F^{r}_{\nu}}$ and the neutrino energy $\varepsilon$.
Both the dipole amplitude and the direction of this dipole were observed to
exhibit basically stationary or slowly migrating behavior over time intervals 
of at least several hundred milliseconds covered by the simulations.

In contrast, \citet{2015ApJ...800...10D} did not observe any evidence for 
LESA when carrying out 2D CCSN simulations with a multi-group,
flux-limited diffusion scheme for neutrino transport.
For this reason, \citet{2015ApJS..216....5S} speculated that LESA may be an 
artifact of the RbR+ approximation.
In a later study, \citet{2018MNRAS.477.3091V} also did not find indications of LESA
when conducting 2D simulations with a fully multi-dimensional two-moment scheme.
However, \citet{2014ApJ...792...96T} argued that the behavior of LESA was
considerably different between their 2D and 3D models. While in the 2D case
the LESA dipole direction, which can only be aligned with the artificial axis
of symmetry, flips between both polar directions on a time scale of some 
10\,ms, strongly affected by SASI mass motions around the proto-neutron star,
only their 3D models showed a persistent and directionally stable LESA
dipole.

\citet{2016ApJ...831...98R} also investigated their 3D simulations with a fully
multi-dimensional two-moment neutrino-transport scheme for effects of LESA,
but could not trace any evidence of this phenomenon. Although their models
did not take into account velocity-dependent terms in the neutrino-moment 
equations and ignored inelastic neutrino scattering on electrons, this
group interpreted the absence of LESA in contrast to the SN models of
\citet{2014ApJ...792...96T} not necessarily as a consequence of these
differences in the neutrino treatment. Instead, \citet{2016ApJ...831...98R}
hypothesized that it could be linked, for example, to the fact that neutron-star
convection showed up rather late in their simulations.

The first independent confirmation of the LESA phenomenon in 3D SN models 
with detailed neutrino transport different from those of the Garching group
has recently been reported by \citet{2018ApJ...865...81O} and has possibly
be seen in a 3D simulation by \citet{2019MNRAS.482..351V}, too. Both groups
also applied a fully multi-dimensional two-moment scheme, for which reason
their results provide support for LESA not being an artifact of the RbR
transport approximation. Although \citet{2018ApJ...865...81O} observed
a substantial LESA dipole
only in their single full-sphere calculation including velocity-dependent
terms in the transport, they argued that the appearance of 
LESA is not directly linked to this improvement but appears to be a
consequence of the stronger proto-neutron star convection seen in the
model with velocity dependence. This conclusion may offer an explanation
why LESA was absent in the models of \citet{2016ApJ...831...98R}, and it
is consistent with the
finding by \citet{2016ARNPS..66..341J} that LESA is weaker in rotating
and further diminished in rapidly rotating models, because
angular momentum gradients in the neutron star suppress convective activity.

Here we present an evaluation of a set of eight new 3D simulations published in
a companion paper by \citet[Paper~I]{paper1} for the LESA phenomenon.
The simulations were performed with the \textsc{Aenus-Alcar} code for an exploding 
9\,$M_\odot$ progenitor and a nonexploding 20\,$M_\odot$ star with two chosen 
grid resolutions. The transport treatment for all three neutrino flavors
was based on an energy- and velocity-dependent two-moment scheme (combining
energy and momentum equations with an analytic ``M1'' Eddington closure 
relation), which was applied in a fully multi-dimensional (FMD) version as
well as an RbR+ mode. In all of these eight simulations (four for the 
low-mass and four for the high-mass progenitor), we find clear evidence for
the LESA with all of its characteristic properties described by 
\citet{2014ApJ...792...96T} and \citet{2014PhRvD..90d5032T}.

Our paper is structured as follows. In Section~\ref{sec:numerics} we briefly
summarize the numerical methods and computed models, refering the reader for
more detailed information to Paper~I. In Section~\ref{sec:lesa} we present the
results from our LESA analysis of the 3D simulations with FMD and RbR+ 
transport treatments. In Section~\ref{sec:LESA-PNS} we provide a more 
detailed description of the proto-neutron star convection layer in the 
phase where LESA with a dominant dipole mode is fully developed. 
In Section~\ref{sec:LESA-Chandra} we construct a tentative link between
the growth and initial evolution of different LESA multipoles and the
onset of thermally driven convection discussed via a linear stability
theory by \citet{1961hhs..book.....C}. In Section~\ref{sec:conclusion}
we wrap up by an assessment of our results and conclusions.

\section{Numerics, Simulation Setup,\\and 3D Models}
\label{sec:numerics}

This study of LESA is based on the 3D simulations that were performed in Paper I.
We will briefly summarize the numerical methods and the simulation setup in the following,
and refer the reader to Paper I for detailed descriptions.

\begin{deluxetable*}{lclccc}
    \tablecaption{ Overview of 3D simulations.\label{tab:simulations}}
    \tablewidth{0pt}
    \tablehead{
        \multicolumn{1}{c}{Model Name} &
        \multicolumn{1}{c}{Progenitor} &
        \multicolumn{1}{c}{Transport} &
        \colhead{Resolution} &
        \colhead{$N_{r} \times N_{\theta}$ $\times N_{\phi}$} &
        \colhead{Explosion}
    }
    \startdata
    s20 FMD L   &  s20   & FMD  & L  & $320 \times 40 \times 80$   & no  \\
    s20 RbR+ L  &  s20   & RbR+ & L  & $320 \times 40 \times 80$   & no  \\
    s20 FMD H   &  s20   & FMD  & H  & $640 \times 80 \times 160$  & no  \\
    s20 RbR+ H  &  s20   & RbR+ & H  & $640 \times 80 \times 160$  & no  \\\hline
    s9.0 FMD L  &  s9.0  & FMD  & L  & $320 \times 40 \times 80$  & yes \\
    s9.0 RbR+ L &  s9.0  & RbR+ & L  & $320 \times 40 \times 80$  & yes \\
    s9.0 FMD H  &  s9.0  & FMD  & H  & $640 \times 80 \times 160$ & yes \\
    s9.0 RbR+ H &  s9.0  & RbR+ & H  & $640 \times 80 \times 160$ & yes \\
    \enddata
\end{deluxetable*}

We employ the \textsc{Aenus-Alcar} code \citep{2008ObergaulingerPhD,2015MNRAS.453.3386J,2018MNRAS.481.4786J},
which solves the hydrodynamics equations and the velocity- and energy-dependent neutrino-moment equations
in spherical polar coordinates (given by radius $r$, polar angle $\theta$, and azimuth angle $\phi$)
with a Godunov-type, directionally unsplit finite-volume scheme
and high-order reconstruction methods with an approximate Riemann solver to obtain cell-interface fluxes.
We make use of the microphysical equation of state (SFHo) from \citet[][]{2013ApJ...774...17S}
to close the system of hydrodynamics equations
and include spherically symmetric gravitational self-interaction
with general relativistic corrections \citep[see][case A]{2006A&A...445..273M}.
In the ``M1'' two-moment neutrino-transport scheme the energy density ${E_{\nu}}$
and energy-flux density ${F^{i}_{\nu}}$ (with ${i \in \{r,\theta,\phi\}}$)
of neutrinos $\nu$ are evolved for electron neutrinos \nue, anti-electron neutrinos \nubar,
and a third species \nux that represents all four heavy-lepton neutrinos.
The system of moment equations is closed by an algebraic relation for non-evolved moments like the radiation pressure tensor
that depends on ${E_{\nu}}$ and ${F^{i}_{\nu}}$.
In contrast to the FMD two-moment transport scheme,
the non-radial neutrino-flux components are set to zero in calculations with the RbR+ approximation.
In the 3D simulations discussed here, we used 15 bins for the neutrino energy $\varepsilon$,
that are logarithmically spaced in the interval ${0 \le \varepsilon \le 400\unitspace\mathrm{MeV}}$
for all neutrino species.
The employed neutrino interactions are specified in Paper I.

In total, we conducted eight self-consistent, time-dependent CCSN simulations in 3D
for two different non-rotating, solar-metallicity progenitor models (see Table~\ref{tab:simulations}):
the ${20\unitspace\msun}$ progenitor model from \citet[][labeled s20 in the following]{2007PhR...442..269W},
and the ${9\unitspace\msun}$ solar-metallicity progenitor model from \citet[][labeled s9.0 in the following]{2015ApJ...810...34W}.
For simulations of both progenitor models, we changed the numerical grid resolution by varying the number of grid cells
($N_{r}$, $N_{\theta}$, and $N_{\phi}$, respectively).
Simulations with ``low'' (``L'') resolution use ${N_{r} = 320}$, ${N_{\theta} = 40}$, and ${N_{\phi} = 80}$ zones,
whereas models with ``high'' (``H'') resolution use twice the number of cells, i.e.,
${N_{r} = 640}$, ${N_{\theta} = 80}$, and ${N_{\phi} = 160}$.
In all cases, the radial grid ranges from 0 to ${10^{9}\unitspace\textrm{cm}}$ with logarithmically increasing cell sizes,
and the angular grids cover the entire solid angle of the full sphere
with cell sizes of ${4\unitspace\degree - 4.5\unitspace\degree}$ (low) and ${2\degree - 2.25\degree}$ (high)
except for two bigger lateral zones at each of both poles of the spherical coordinate system (for details, see Paper I).
In contrast, the azimuthal grid is equidistant.
For each progenitor model and grid resolution, we performed simulations with the RbR+ approximation and the FMD neutrino-transport scheme.

The small cell sizes near the center of the polar coordinate system lead to severe constraints on the numerical time step,
which we circumvent by evolving the innermost ${10^{6}\unitspace\mathrm{cm}}$ in spherical symmetry.
Thus, we neglect angular variations in the (spherically symmetric) innermost region of the proto-neutron star.
This is an acceptable approach as long as the central core of ${r < 10^{6}\unitspace\mathrm{cm}}$ remains convectively stable.
Practically this is only fulfilled until ${\sim100\unitspace\mathrm{ms}}$ after bounce
but can become problematic in some of our simulations at later times
(see end of Section~\ref{sec:LESA-PNS} and Section~\ref{sec:LESA-Chandra}).

\section{LESA in FMD and RbR+ Models}
\label{sec:lesa}

In order to shed more light on the rather unclear situation pictured by
the diverse (but not necessarily conflicting) results of the presence or absence of LESA in models reported in the literature,
we also evaluate our 3D simulations for signatures of LESA.
We can exploit the unique possibility to directly \new{assess} the influence of the neutrino-transport schemes, FMD vs.\
RbR+. To this end we first perform a multipole analysis and decompose the 
radial lepton-number flux
in the laboratory frame (i.e., the rest frame of the stellar center),
\begin{equation}\label{eq:lnf}
F_{\mathrm{lnf}}^{\mathrm{lab}} =
F^{r,\mathrm{n}}_{\nu_{e},\mathrm{lab}} -
F^{r,\mathrm{n}}_{\bar{\nu}_{e},\mathrm{lab}},
\end{equation}
into real spherical harmonics, ${Y^{m}_{l}}$, of degree $l$ and order $m$.
The radial number fluxes of neutrinos in the lab frame are obtained by the transformation
\begin{equation}\label{eqn:flux_lab_frame}
F^{r,\mathrm{n}}_{\nu,\mathrm{lab}} = F^{r,\mathrm{n}}_{\nu}+v_{r}N_{\nu}\,,
\end{equation}
where $v_r$ is the radial fluid velocity and 
$N_{\nu} = \int E_{\nu} \; \varepsilon^{-1} \dint{\varepsilon}$ 
is the number density of neutrinos in the co-moving frame.
The multipole coefficients at a given radius $r$ are calculated by integrating 
the flux over the whole spherical surface:
\begin{equation}\label{eqn:lnf_coefficients}
c^{m}_{l} (r) =
\sqrt{\frac{4 \pi}{2 l + 1}}
\int r^{2}
F_{\mathrm{lnf}}^{\mathrm{lab}}
Y^{m}_{l} \dint\Omega .
\end{equation}
The moments of the lepton-number flux are then obtained as
\begin{equation}\label{eqn:lnf_dipole}
A_{\mathrm{lnf}} (r, l) = \sqrt{\sum_{m=-l}^{l} \left[ c^{m}_{l} (r) \right] ^{2}} .
\end{equation}
Here, we use a normalization for which the monopole moment, ${A_{\mathrm{lnf}} (r, 0)}$,
is identified with the total lepton-number flux through the surface of a sphere with radius $r$\footnote{We point out that
the normalization of the spherical harmonics coefficients in the present case differs
from what we used in the decomposition of the shock radius in Equations~(11) and (12) of Paper I.
There, all coefficients $R_{\mathrm{s}}^{l}$ are rescaled by a factor of $1 / (4 \pi)$ compared to $A_{\mathrm{lnf}} (r, l)$.
With this normalization the monopole moment stands for the mean shock radius.}.

\begin{figure}
    \includegraphics[width=0.48\textwidth]{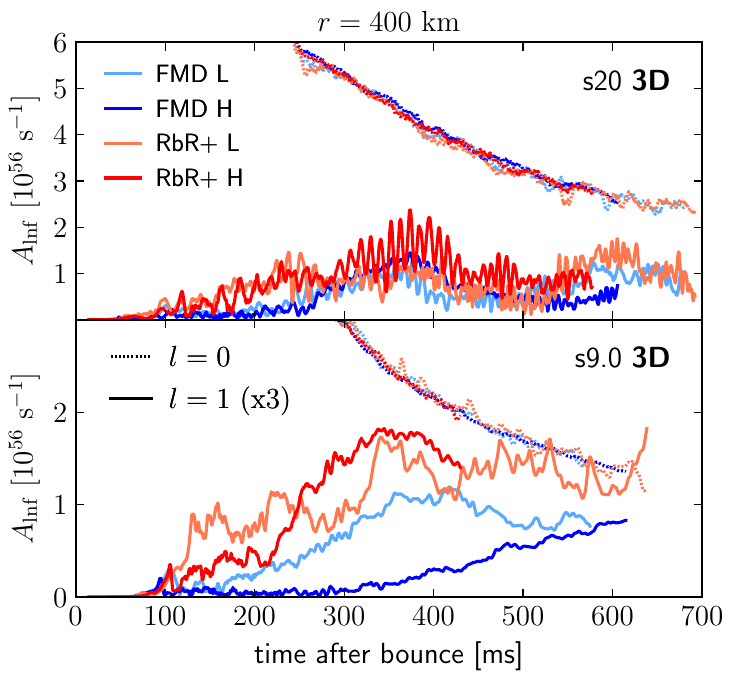}
    \caption{Monopole (${l=0}$, dotted lines) and dipole ($l=1$, solid lines) moments
            of the electron lepton-number flux, evaluated at a radius of 
            ${r = 400\unitspace\mathrm{km}}$ and transformed into the lab frame 
            at infinity,
            as functions of time after core bounce for all 3D simulations
            of the s20 (top panel) and the s9.0 (bottom panel) progenitor models.
            The multipole amplitudes $A_{\mathrm{lnf}}$ are defined as given in 
            Equation~\eqref{eqn:lnf_dipole}. While for ${l=0}$ the corresponding 
            amplitude agrees with the total electron lepton-number flux, the 
            plot shows the dipole amplitude of Equation~\eqref{eqn:lnf_dipole}
            scaled by a factor of 3 for consistency with the normalization 
            used by \citet{2014ApJ...792...96T}: $A_0 + A_1\cos\vartheta$ with
            $A_0$ and $A_1$ being the monopole and dipole amplitudes, respectively,
            and $\vartheta$ the angle relative to the dipole direction.
            \label{fig:lesa_multipoles_s20_s90}}
\end{figure}

\begin{figure*}
    \includegraphics[width=\textwidth]{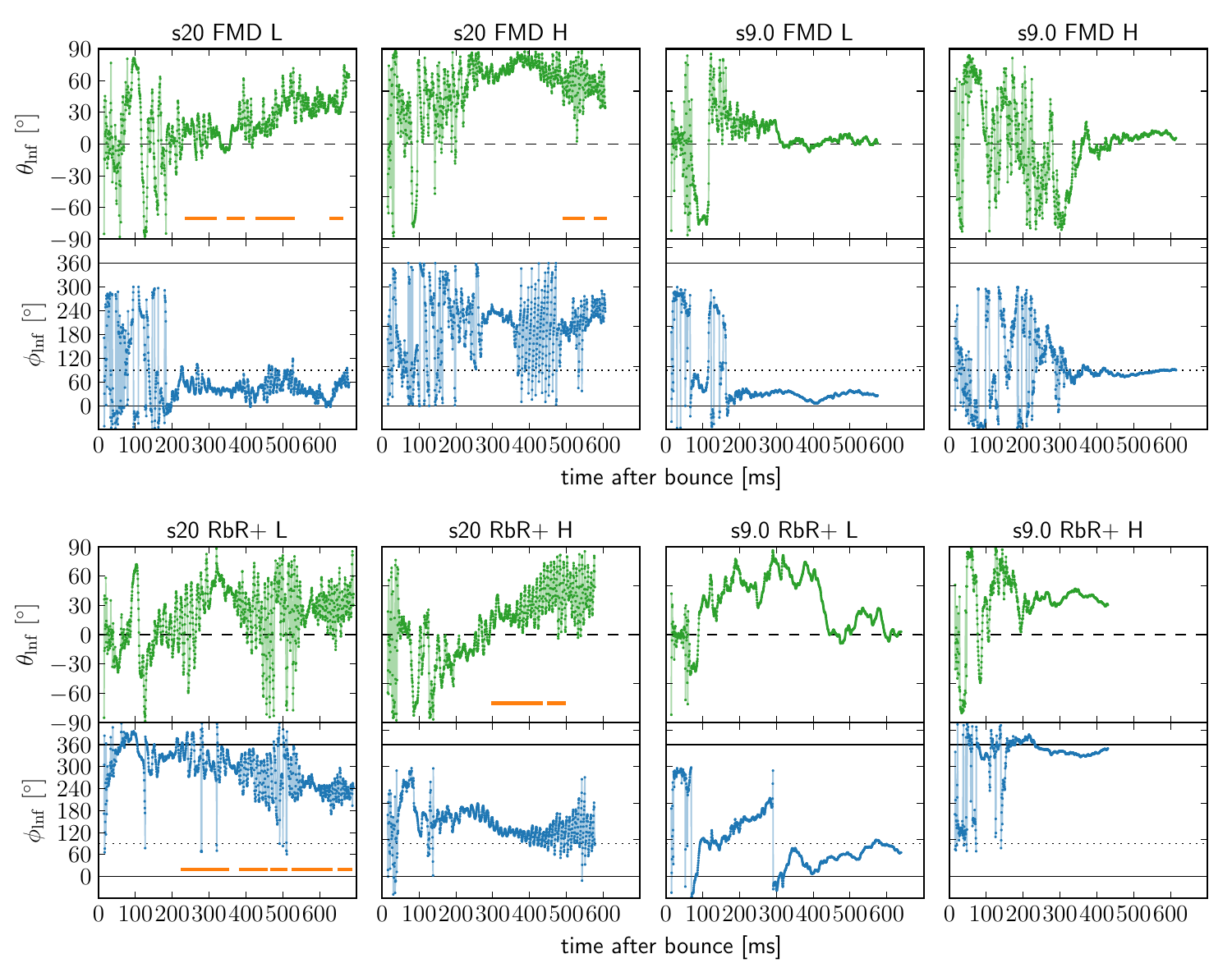}
    \caption{Time evolution of the direction of the electron lepton-number flux dipole
            (i.e., of the amplitude of the ${l=1}$ moment plotted in Figure~\ref{fig:lesa_multipoles_s20_s90})
            for all 3D simulations in terms of
            the polar angle ${\theta_{\mathrm{lnf}}}$ (green dots \new{connected by shaded lines})
            and the azimuthal angle ${\phi_{\mathrm{lnf}}}$ (blue dots \new{connected by shaded lines}) of the dipole vector.
            The upper row of panels shows the FMD runs with low and high
            resolution for the s20 and s9.0 progenitors, the lower row
            the corresponding RbR+ results.
            \new{For better readability the values of the polar angle
            are transformed to range within the interval
            ${\theta_{\mathrm{lnf}} \in [-90^{\circ}, +90^{\circ}]}$, which implies
            that the equator is located at ${\theta_{\mathrm{lnf}} = 0^{\circ}}$ (dashed black line).
            Moreover, some values of the periodic azimuthal angle are shifted by adding or subtracting $360^{\circ}$ in some panels.
            The $x$ and $y$ directions of an associated Cartesian coordinate
            system correspond to ${\phi_{\mathrm{lnf}} = 0^{\circ}}$ (or, equivalently, ${\phi_{\mathrm{lnf}} = 360^{\circ}}$, solid black lines)
            and ${\phi_{\mathrm{lnf}} = 90^{\circ}}$ (dotted black lines), respectively.}
            Note that for directions very close to the poles,
            i.e.\ ${\theta_{\mathrm{lnf}} \sim \pm90^{\circ}}$, variations of the
            azimuthal angle $\phi_\mathrm{lnf}$, e.g.\ in model `s20 FMD H'
            between $\sim$350\,ms and $\sim$500\,ms after bounce, describe migrations
            within a small area around the poles. This is obvious from
            Figure~\ref{fig:lesa_dipole_direction_aitoff_s20_s90}, where the 
            evolution of the dipole directions for all high-resolution simulations
            is visualized in terms of Aitoff projections of the angle space.
            The time intervals marked by orange horizontal bars roughly indicate SASI episodes.
            \label{fig:lesa_direction_s20_s90}}
\end{figure*}

Figure~\ref{fig:lesa_multipoles_s20_s90} displays the time-dependent evolution of the 
amplitudes of the monopole
(${l=0}$, dotted lines) and of the dipole (${l=1}$, solid lines, scaled by a factor
of 3 compared to Equation~\eqref{eqn:lnf_dipole}) of the total electron lepton-number flux
(evaluated as lab-frame quantities at a radius of ${r = 400\unitspace\mathrm{km}}$) 
for all of our 3D simulations. To visualize the evolution of the dipole direction,
the polar and azimuthal angles of the dipole vector\footnote{For
the dipole direction of the lepton-number flux, we use the multipole coefficients 
${c^{m}_{1} (r)}$ as components of the direction vector,
which we then transform into spherical polar coordinates.
This procedure is equivalent to the analysis of the dipole direction of the shock radius
(see Equations~(11) and (12) and the associated explanations in Paper~I).} 
are shown as functions of time after bounce in Figure~\ref{fig:lesa_direction_s20_s90}.

All of our simulations, both with RbR+ and FMD neutrino transport, exhibit dipole 
components that grow with time and reach between \new{$\sim$10\%} and more than \new{30\%}
\new{(keeping in mind the scaling in Figure~\ref{fig:lesa_multipoles_s20_s90} by a factor of 3)}
of the monopole amplitude at the end of the calculations. While the individual
characteristics of the evolution differ from model to model, a clear correlation can
be observed in all cases between the time when the dipole amplitude reaches a 
substantial fraction of the monopole and the time when the direction of the dipole 
approaches a fairly stable state with only slow and modest subsequent migration.
This can be most clearly seen in the SN runs for the s9.0 progenitor, for example
around 200\,ms p.b.\ in models `s9.0 FMD L' and `s9.0 RbR+ H', and at $\sim$350\,ms
in `s9.0 FMH H', whereas in model `s9.0 RbR+ L' the latitudinal
angle reaches a well-defined value after about 100\,ms while the azimuthal angle
continues to drift until $\sim$350\,ms p.b. But this motion takes place in a rather
small region around the pole of the grid since $\theta_\mathrm{lnf}\gtrsim +70^\circ$
(compare Figures~\ref{fig:lesa_multipoles_s20_s90} and
\ref{fig:lesa_direction_s20_s90}). 

\begin{figure*}
    \includegraphics[width=\textwidth]{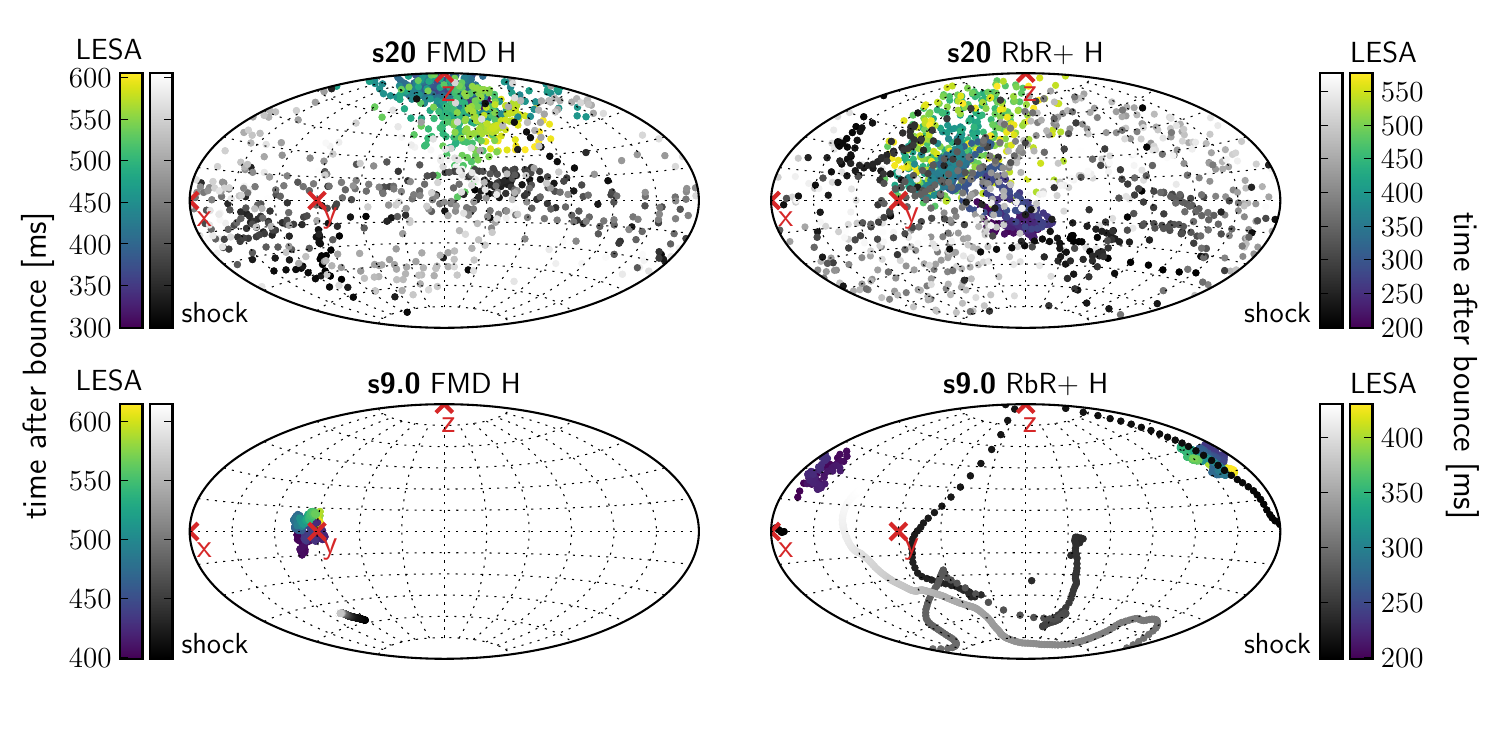}
    \caption{Time evolution of the directions of the electron lepton-number flux (LESA) dipole
            and of the dipole vector of the deformed shock surface for all high-resolution
            3D simulations visualized by Aitoff projections
            of the angle space. In the case of the s20 progenitor the shock dipole mainly
            corresponds to an asphericity of the shock caused by violent SASI sloshing and 
            spiral motions, which also lead to direction-dependent modulations of the 
            neutrino luminosity due to asymmetric accretion of the neutron star.
            The dots mark the location of the dipoles at certain times with the colors from
            blue to yellow for the LESA dipole and dark gray to light gray for the shock
            dipole indicating the direction of time (see color bars). 
            The time intervals of the displayed
            evolution differs between the different panels, covering in each case the 
            phase of the most pronounced LESA dipole. The axis directions $x$, $y$, and $z$
            of the Cartesian coordinate system, underlying the definition of the direction
            angles used in Figure~\ref{fig:lesa_direction_s20_s90}, are marked by red crosses.
            In the convection-dominated s9.0 models (bottom panels) SASI plays no role. The
            dipole deformation mode of the shock is caused by asymmetric convection and has
            hardly any influence on the LESA dipole. In model `s20 FMD H' 
            (upper left panel) SASI mass motions and accretion asymmetries occur in a 
            plane roughly perpendicular to the LESA dipole, whose amplitude
            and orientation are therefore affected less strongly by the SASI-induced 
            neutrino-emission modulations than in model `s20 RbR+ H' \new{(upper right panel)},
            where the LESA and SASI dipole directions coincide temporarily.
            \new{The alignment of the dipole direction with the y-axis in model `s9.0 FMD H' (lower left panel)
            is just accidental, because the transformation from spherical polar coordinates to
            Cartesian coordinates in the $x$-$y$-plane is arbitrary.}
            \label{fig:lesa_dipole_direction_aitoff_s20_s90}}
\end{figure*}

In the s20 runs the dipole direction also has the tendency to evolve more slowly
after an initial phase of basically random motion during the period when the
amplitude still grows. However, different from the convection-dominated 
post-shock layer in the s9.0 models, the post-bounce dynamics of the s20 runs is
characterized by repeated episodes of violent SASI activity. As shown by 
\citet{2014ApJ...792...96T}, the LESA dipole is mostly formed in the convective
layer interior of the proto-neutron star and further enhanced by the lepton-number loss
from the accretion layer around the neutrinosphere, where the electron fraction
develops a pronounced hemispheric asymmetry.
For this reason \citet{2014ApJ...792...96T}, \citet{2014PhRvD..90d5032T}, \new{and \citet{2019arXiv190106235W}}
observed that SASI-induced variations of the accretion
flow between stalled shock and neutron star and the associated time- and 
direction-dependent modulations of the accretion luminosity (mostly $\nu_e$ 
plus $\bar\nu_e$) can have a considerable impact on the evolution of the LESA dipole.
This influence of SASI on LESA is strongest when the LESA direction coincides
with the plane or even the direction of the SASI sloshing or spiral motions.
\new{Regarding the interplay between SASI and LESA, we do not witness any obvious dependence on the employed transport method.}

\begin{figure*}
    \includegraphics[width=\textwidth]{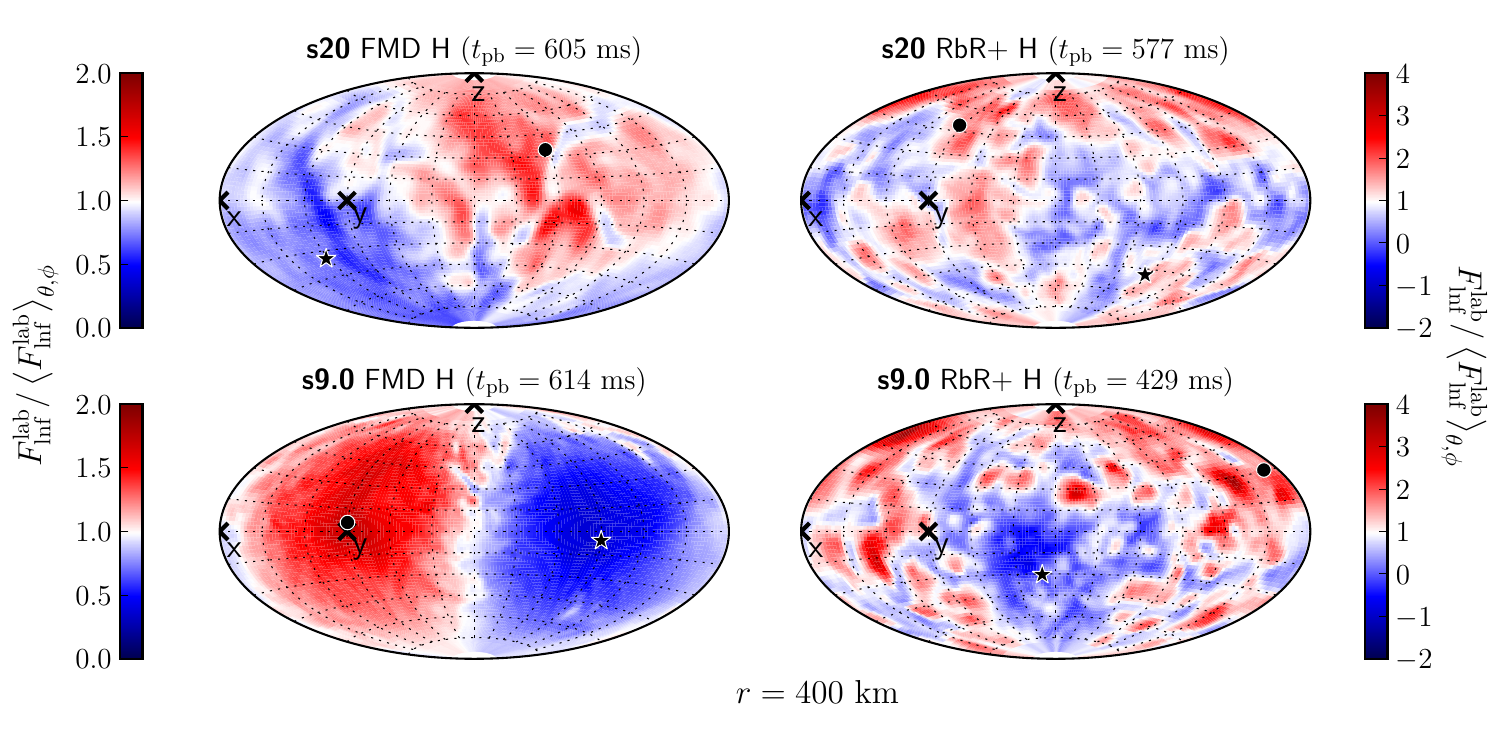}
    \caption{Aitoff projections of the electron lepton-number flux,
            $F_{\mathrm{lnf}}^{\mathrm{lab}}$ (see Equation~\eqref{eq:lnf}),
            evaluated at a radius of ${r = 400\unitspace\mathrm{km}}$, transformed
            into the lab frame at infinity, and normalized by the angular average,
            ${\left< F_{\mathrm{lnf}}^{\mathrm{lab}} \right>_{\theta, \phi}}$.
            The results are displayed for all high-resolution 3D runs (arrangement of
            panels analogous to Figure~\ref{fig:lesa_dipole_direction_aitoff_s20_s90})
            at the end of each simulation (given by $t_{\mathrm{pb}}$ in the figure labels).
            The directions of the LESA dipole moments (see final results in
            Figures~\ref{fig:lesa_direction_s20_s90} and \ref{fig:lesa_dipole_direction_aitoff_s20_s90})
            are marked by circles, the opposite directions by stars.
            The dominance of a dipole asymmetry is visible in all cases, but the spottiness of the RbR+
            results suggests more important contributions also from higher-order multipoles. This
            is confirmed by the ``spectrograms'' shown in Figure~\ref{fig:lesa_multipoles_cmap_s20_s90}.
            \new{The exceptionally clear hemispheric difference in model `s9.0 FMD H'
            (and the very stable dipole direction seen in Figure~\ref{fig:lesa_dipole_direction_aitoff_s20_s90})
            result from (1) the absence of SASI and the low mass-accretion rate in downflows in this exploding model,
            (2) the ``smearing'' of angular variations by the nonradial flux components that are evolved in the FMD scheme,
            but neglected in the RbR+ transport,
            and (3) the fact that `s9.0 FMD H' is shown at a time of evolution when the LESA dipole mode is largest
            and strongly dominates the higher multipoles (see Figures~\ref{fig:lesa_multipoles_s20_s90} and \ref{fig:lesa_multipoles_cmap_s20_s90}).}
            For orientation, also the $x$, $y$, and $z$ directions of the Cartesian coordinate
            system for the measurement of the dipole angles of Figure~\ref{fig:lesa_direction_s20_s90}
            are indicated by black crosses.
            \label{fig:lesa_flux_aitoff_s20_s90}}
\end{figure*}

\begin{figure*}
    \includegraphics[width=\textwidth]{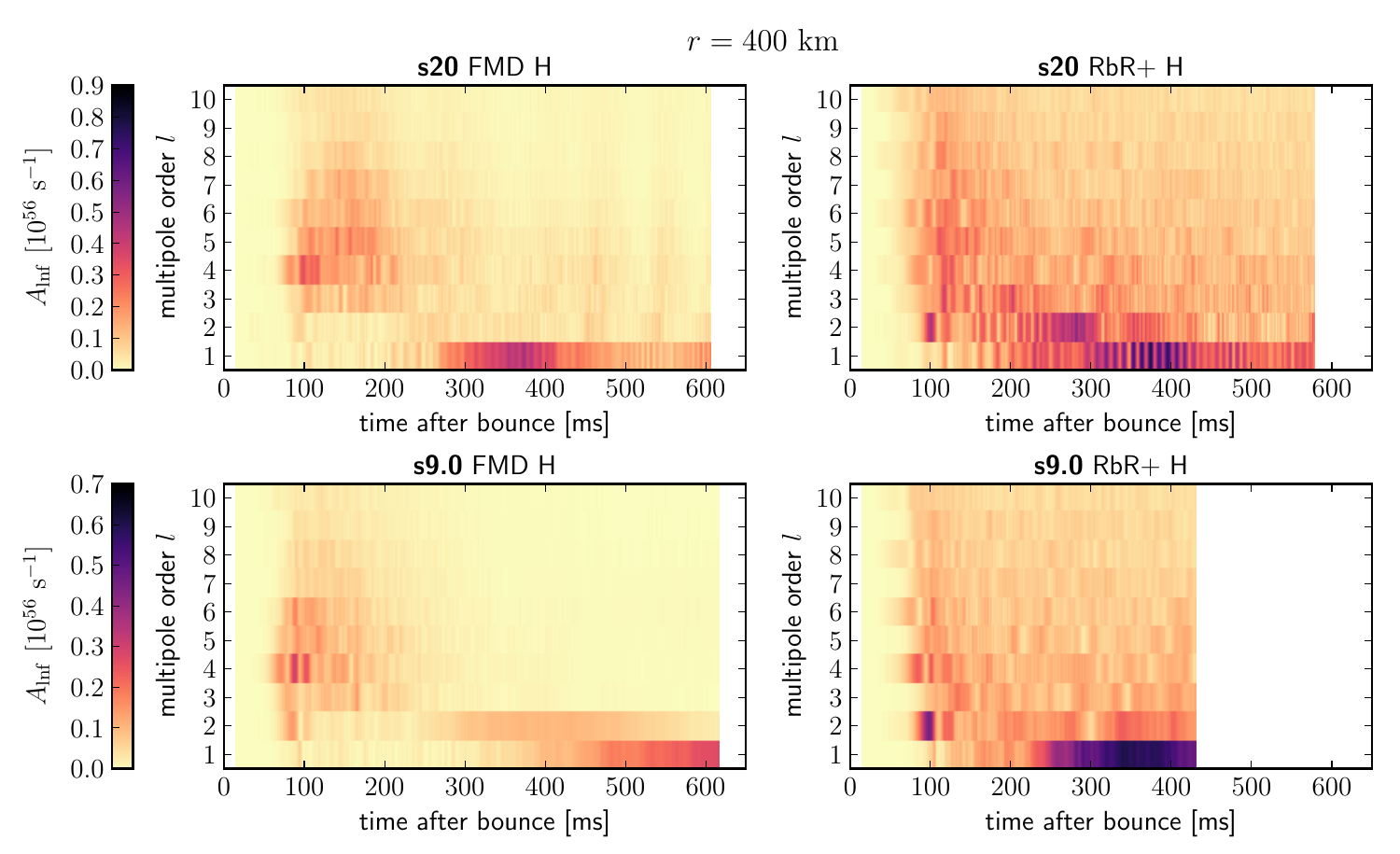}
    \caption{``Spectrograms'' for the time evolution of the multipole moments $A_{\mathrm{lnf}}$
            of the electron lepton-number flux (as defined in Equation~\eqref{eqn:lnf_dipole})
            for orders $l = 1$ to $l = 10$ (evaluated at a radius of ${r = 400\unitspace\mathrm{km}}$
            for an observer in the lab frame at infinity). Results from all of our four
            high-resolution 3D simulations are displayed (arranged analogoue to the panels in
            Figures~\ref{fig:lesa_dipole_direction_aitoff_s20_s90} and
            \ref{fig:lesa_flux_aitoff_s20_s90}). The color intensity scales with the
            multipole amplitude (see color bars). To optimize the visualization of the
            relative strengths of the moments for $l \ge 1$, the monopole (${l = 0}$),
            which would far dominate the early post-bounce evolution, is not displayed.
            \label{fig:lesa_multipoles_cmap_s20_s90}}
\end{figure*}

The corresponding effects can be clearly spotted in the s20 runs, see 
Figures~\ref{fig:lesa_multipoles_s20_s90} (upper panel) and \ref{fig:lesa_direction_s20_s90},
where phases exist when both the dipole amplitude and the
direction angles exhibit pronounced oscillations. These oscillations correlate
with episodes of strong SASI activity in the post-shock layer 
(see Section~4.1 and Figure~6 of Paper~I). The overlap
of SASI and LESA directions is particularly evident in model `s20 RbR+ H',
where between about 200\,ms and 450\,ms after bounce the LESA dipole and the
SASI direction are roughly aligned as visible in the upper right panel of
Figure~\ref{fig:lesa_dipole_direction_aitoff_s20_s90}. Exactly during this
time interval the LESA dipole amplitude (Figure~\ref{fig:lesa_multipoles_s20_s90})
shows large quasi-periodic modulations and both direction angles of the LESA
dipole exhibit a major shift by $\sim$90$^\circ$ instead of 
the usual temporal fluctuations, which seem to be a consequence
of the time- and directional variability of the massive accretion flows to the
neutron star in the nonexploding s20 models (Figure~\ref{fig:lesa_direction_s20_s90}).
While the ubiquitous temporal fluctuations of the direction angles are also 
present in `s20 FMD H', this case does not display any quasi-monotonic
large-scale migration of 
the LESA direction at $t_\mathrm{pb} \gtrsim 250$\,ms, except for local motions
around the pole at $\theta_\mathrm{lnf} = 90^\circ$, signalled by large 
excursions of $\phi_\mathrm{lnf}$ between $\sim$350\,ms and $\sim$500\,ms
after bounce. This is compatible with the fact that in model `s20 FMD H'
the LESA dipole direction is far off the main plane of SASI activity 
during most of the time (see
upper left panel in Figure~\ref{fig:lesa_dipole_direction_aitoff_s20_s90}).
For the Aitoff projections of the SASI direction in the latter figure we
consider the dipole vector associated with the shock 
deformation\footnote{We use the multipole coefficients $a_x = a_1^1$, 
$a_y = a_1^{-1}$, and $a_z = a_1^0$ from the decomposition
of the angle-dependent shock radius into real spherical harmonics
(see Equations~(11) and (12) and the associated explanations in Paper~I)
as components of the direction vector, which we then transform into 
spherical polar coordinates.}, following \citet{2014ApJ...792...96T}.
Alternatively, \citet{2014PhRvD..90d5032T} employed the dipole of the
neutrino-energy flux, summed over all six neutrino species. The spatial
motions of both direction vectors are closely correlated, and both can
therefore be used equally well as tracers of the SASI motions and of the
associated accretion-modulated anisotropic neutrino emission, which then 
has an impact on the LESA dipole.

In contrast, the runs of the s9.0 progenitor explode rather early 
($t_\mathrm{pb}\sim 300$\,ms) and do not possess any preceding SASI
activity. Moreover, convective fluctuations of the mass-accretion rate 
also have a much reduced influence on the neutrino emission in these 
simulations because of the low mass-accretion rate of this low-mass progenitor.
For these reasons the LESA amplitude (Figure~\ref{fig:lesa_multipoles_s20_s90},
lower panel) and direction angles (Figure~\ref{fig:lesa_direction_s20_s90},
two right columns and Figure~\ref{fig:lesa_dipole_direction_aitoff_s20_s90},
bottom panels) do not show any
phases of quasiperiodic oscillations as in the s20 models, and the LESA 
vector is more stationary or shifts only slowly on longer timescales.

Figure~\ref{fig:lesa_flux_aitoff_s20_s90} presents the variations of the
local electron lepton-number flux densities in all emission directions
for all high-resolution 3D simulations at the times when the runs were 
terminated. The dominance of a dipolar mode is visible in all cases, but 
also contributions from higher multipolar models can be recognized.
This is confirmed by Figure~\ref{fig:lesa_multipoles_cmap_s20_s90},
which provides ``spectrograms'' for the temporal evolution of the moments 
$A_{\mathrm{lnf}}$ from order ${l=1}$ up to ${l = 10}$ for the same set
of simulations. Both of the models with FMD transport exhibit an early
phase between $\sim$${50\unitspace\mathrm{ms}}$ p.b.\ and 
$\sim$${200\unitspace\mathrm{ms}}$ p.b.\ in which only the multipole modes
around ${l = 4}$ are excited. The growth of these asymmetries coincides with
the time when convective activity inside of the proto-neutron star sets in
($t_\mathrm{pb}\sim 40$--50\,ms), whereas nonradial instabilities in the
gain layer start only some 10\,ms later (at $\gtrsim$ 70--80\,ms p.b., see
the nonradial kinetic energies given in Figure~6 of Paper~I).
A dominant dipole begins to appear in
`s20 FMD H' at $t_\mathrm{pb}\gtrsim 250$\,ms, at which time `s9.0 FMH H'
develops most strength in the quadrupole mode, before after $\sim$450\,ms 
of post-bounce evolution also in this model the dipole becomes the clearly
dominant mode, as prominently visible in the lower left panel of
Figure~\ref{fig:lesa_flux_aitoff_s20_s90}. At the end of both runs
the dipole has by far the highest amplitude of all modes with $l\ge 1$.

Also in the simulations with RbR+ transport the modes of higher orders
between $l = 4$ and $l = 6$ appear first, the quadrupole mode reaches 
a dominant amplitude several 10\,ms afterwards, and the dipole mode even
$\sim$150--200\,ms later. And also similar to the FMD runs, as time goes
on there is an obvious tendency that the power in the mode spectrum shifts
to lower and lower modes until the dipole has the highest amplitude.
Different from the FMD runs, for which the lepton-flux asymmetry at
$t_\mathrm{pb}\gtrsim 250$--300\,ms is distinctly concentrated in the
$l=1$ and $l=2$ modes, the RbR+ models possess a broader distribution
of power in the spectrograms. The higher moments, ${l > 1}$, always
yield visible contributions and their power decreases gradually with
higher values of $l$. This is fully compatible with the flux variations
seen in Figure~\ref{fig:lesa_flux_aitoff_s20_s90}, where the RbR+ cases 
exhibit a fragmented pattern of larger and smaller spots (i.e., of lower 
and higher multipole orders), superimposed
on the hemispheric (dipolar) asymmetry. 

This observation, however, is not really astonishing in view of the fact 
that the
RbR+ approximation ignores nonradial components of the neutrino fluxes
and therefore variations in the directions perpendicular to the radius 
vector remain more
localized on smaller scales. In contrast, the nonlocal behavior of
radiative transfer with the FMD scheme, which couples spatial volumes over a neutrino mean free path, leads to a ``smearing'' of angular
variations by the nonradial flux components. This produces much
smoother radiation characteristics around the neutrinosphere and above
it, as discussed in Section~4.1 and as visible in Figures~9 and 10 of Paper~I.
Therefore large-scale or global
directional variations of the neutrino emission such as accretion-induced
hemispheric asymmetries and the LESA phenomenon can be
captured in their basic properties by both FMD and RbR+ transport 
treatments, whereas small-scale variations are stronger with 
RbR+ transport. 

A more detailed comparison reveals further
differences between the results from the two treatments, though
none of these differences is of fundamental nature. For example,
the simulations with RbR+ show a somewhat earlier rise of the LESA dipole 
component for both progenitor models than in the corresponding FMD cases.
Moreover, in the s9.0 models the dipole amplitudes reach somewhat higher 
maximal
values in the RbR+ calculations. Although this does not apply to the
s20 runs, the SASI-induced modulations of the LESA dipole seem to be
more extreme in the RbR+ cases, in particular in model `s20 RbR+ H'
(Figure~\ref{fig:lesa_multipoles_s20_s90}). But these modulations 
depend sensitively on the orientation of the LESA dipole relative to
the SASI direction, which is stochastic and differs from model to model
and can also vary during the post-bounce evolution of a single run. The
migration of the dipole (or its directional stability), although difficult
to compare in detail, does not reveal any qualitatively different behavior
between RbR+ and FMD simulations.

\begin{figure*}[t]
\gridline{\fig{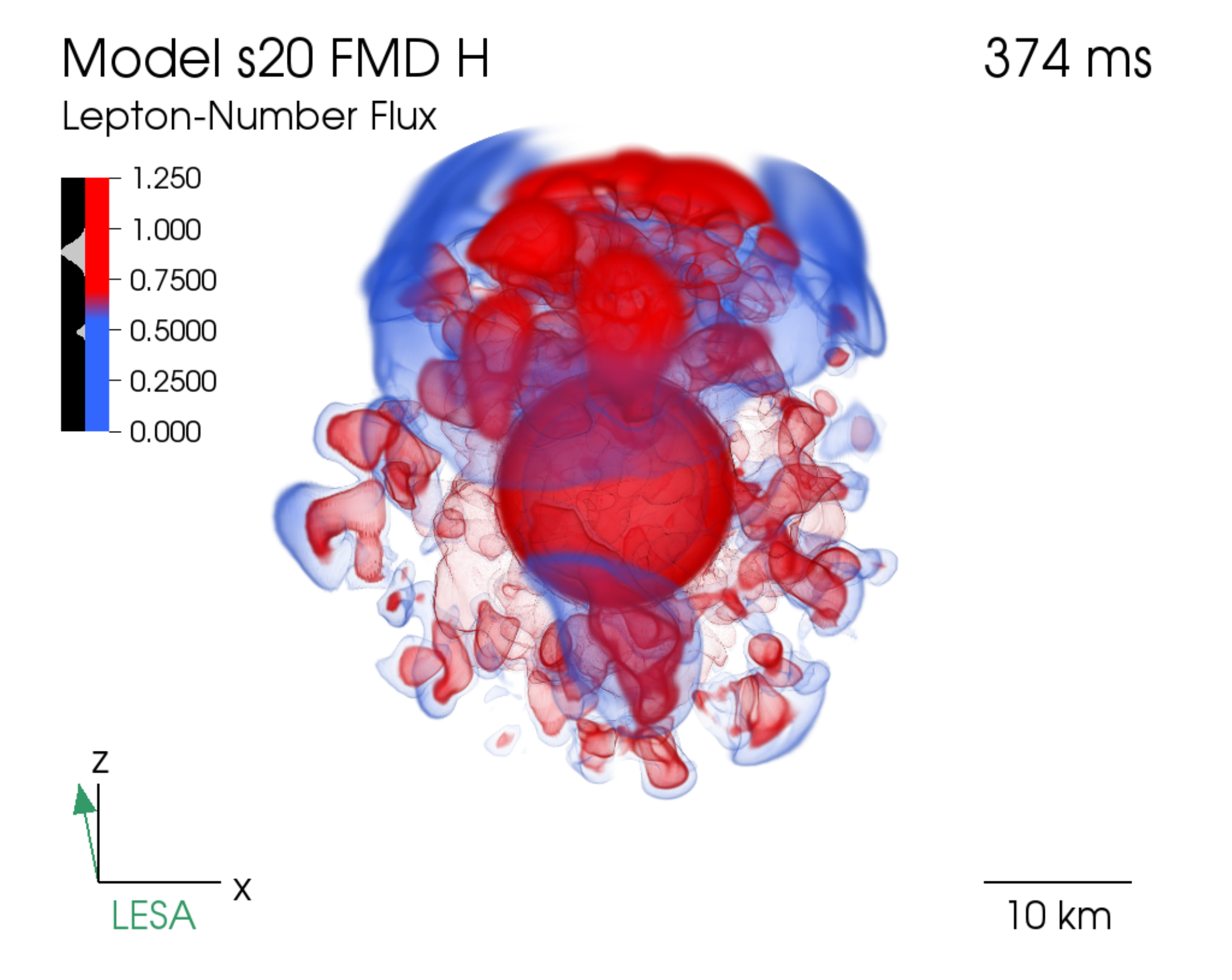}{0.48\textwidth}{}
          \fig{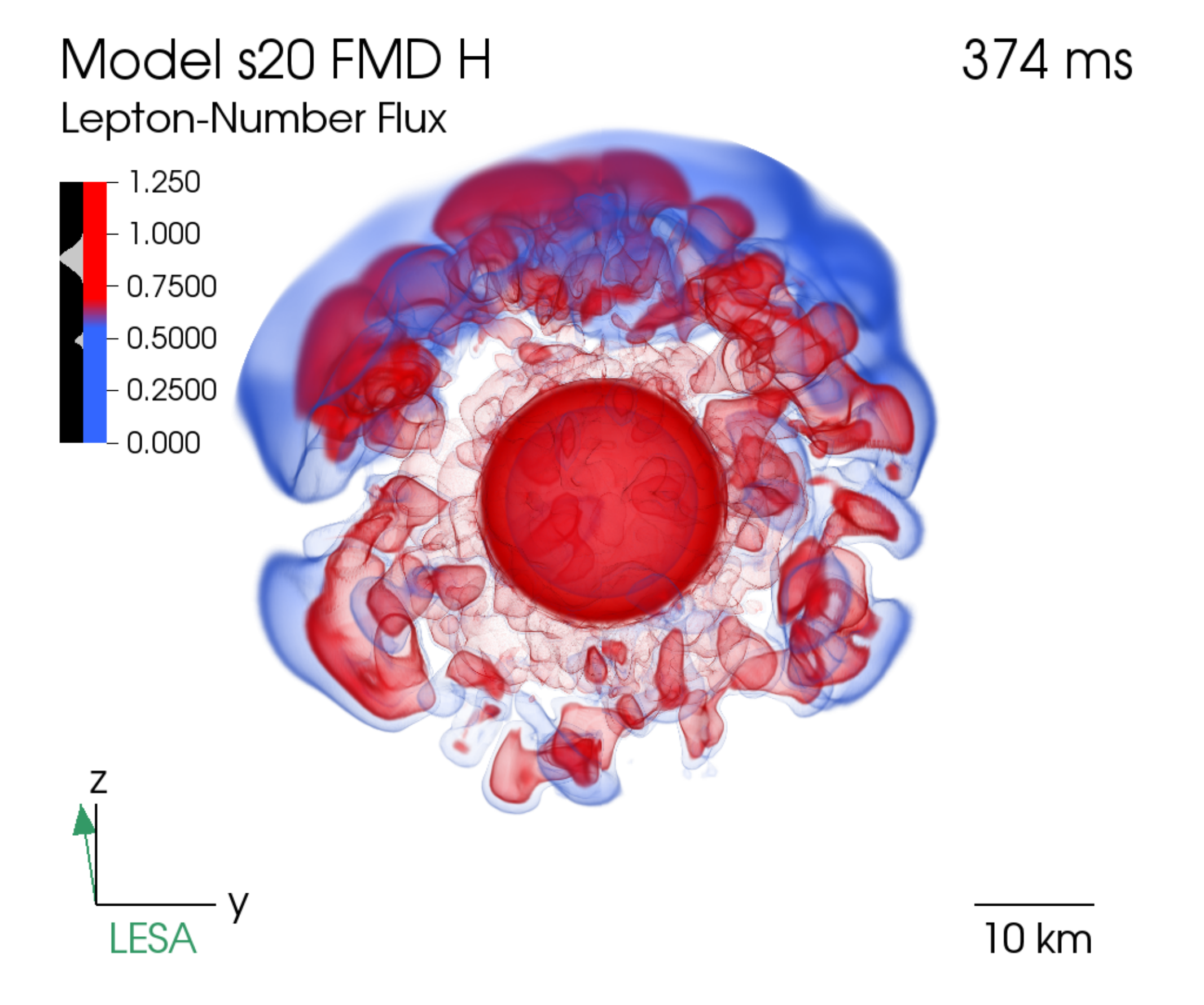}{0.48\textwidth}{}
          }
\gridline{\fig{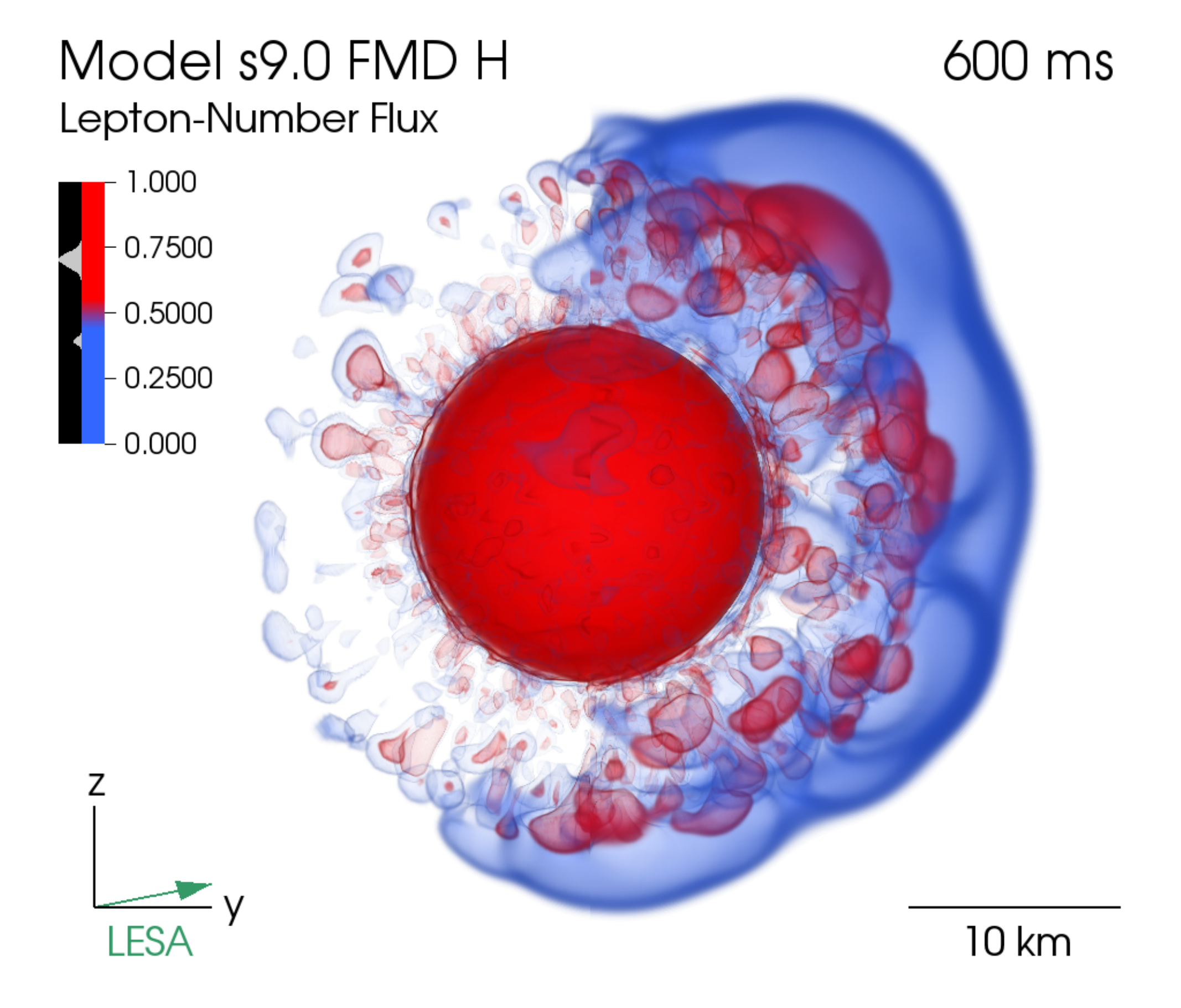}{0.48\textwidth}{}
          \fig{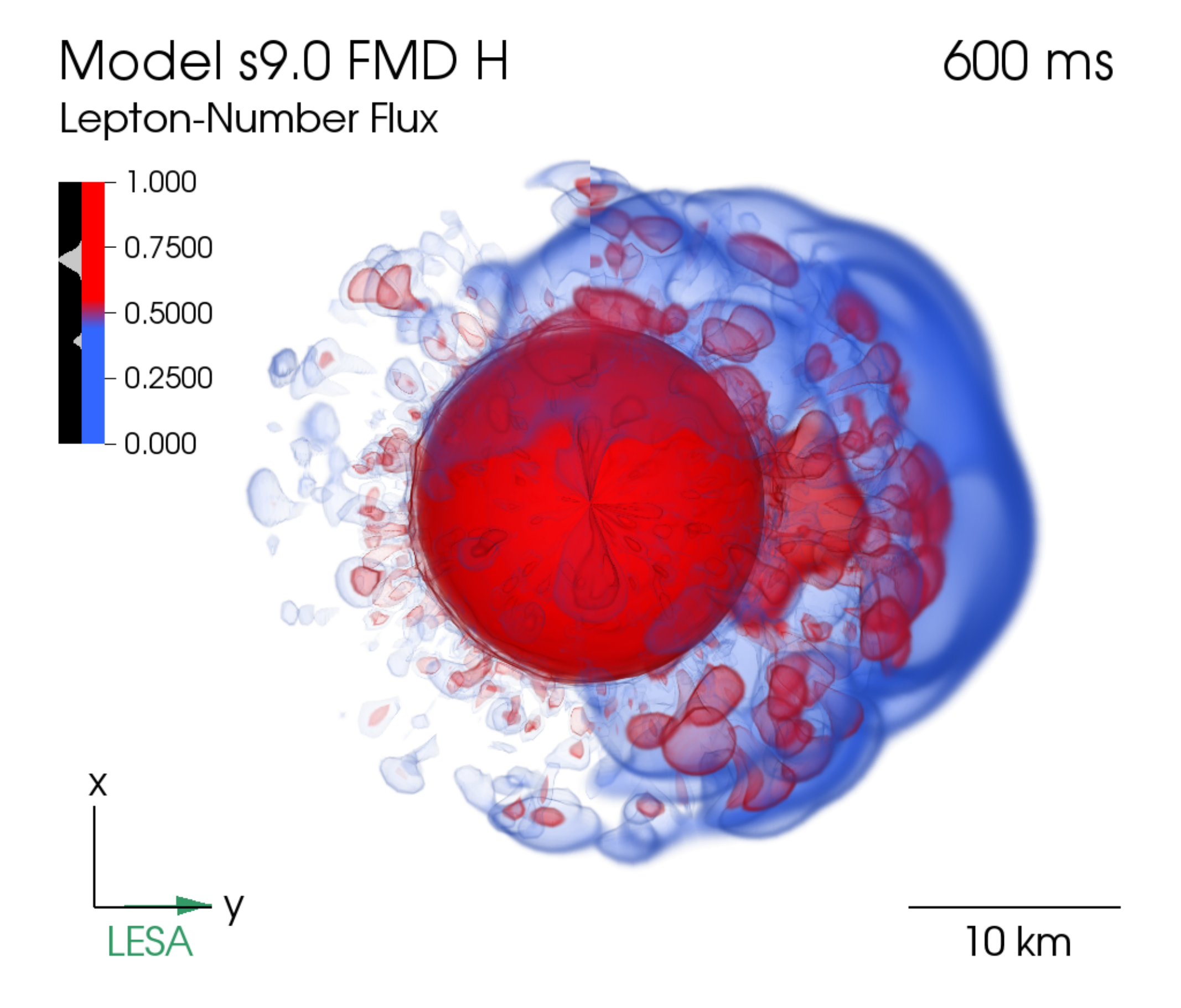}{0.48\textwidth}{}
         }
\caption{\new{Volume renderings of the electron lepton-number flux density in the lab frame,
        $F_{\mathrm{lnf}}^{\mathrm{lab}}$ (see Equations~\eqref{eq:lnf} and \eqref{eqn:flux_lab_frame},
        in units of s$^{-1}$\,cm$^{-2}$ and scaled down by a factor of $10^{43}$),
        in the interior of the neutron star (see yardstick in the lower right corner of each panel).
        For model `s20 FMD H' (top panels) we show data at 374\,ms after bounce,
        and for model `s9.0 FMD H' (bottom panels) we pick a snapshot at 600\,ms, which is close to the end of the simulation.
        At these times, the LESA dipole mode resides near its largest value for each model
        (compare Figure~\ref{fig:lesa_multipoles_s20_s90}).
        The LESA dipole direction, indicated by a green arrow in the tripod in the lower left corner of each panel,
        roughly aligns with the $+z$-direction in model `s20 FMD H',
        and the $+y$-direction in model `s9.0 FMD H' (compare Figure~\ref{fig:lesa_dipole_direction_aitoff_s20_s90}).
        The left and right panels depict different planes, which in all cases contain the LESA vector (see the tripods).
        Since in the neutrino-diffusion regime the visualized lepton-number flux densities are vastly dominated
        by the velocity-dependent term in Equation~\eqref{eqn:flux_lab_frame},
        the images display the hemispheric differences of the small-scale convective cells (visible as red plumes)
        that fill the convection layer in the proto-neutron star.
        The convective cells are much more extended, and convectively enhanced lepton-number transport is much stronger
        (visible by the blue, hemispheric ``caps''),
        on the side of the higher lepton-number emission.
        Opposite to this LESA dipole, convection is weaker and the convective cells are less distinct,
        indicating less efficient convective lepton-number transport.
        Our volume renderings are therefore 3D visualizations of the situation displayed by
        cross-sectional images in Figure 10 of \citet{2014ApJ...792...96T}.}
        \label{fig:lesa_volume_rendering_s20_s90}}
\end{figure*}

\begin{figure*}[t]
\gridline{\fig{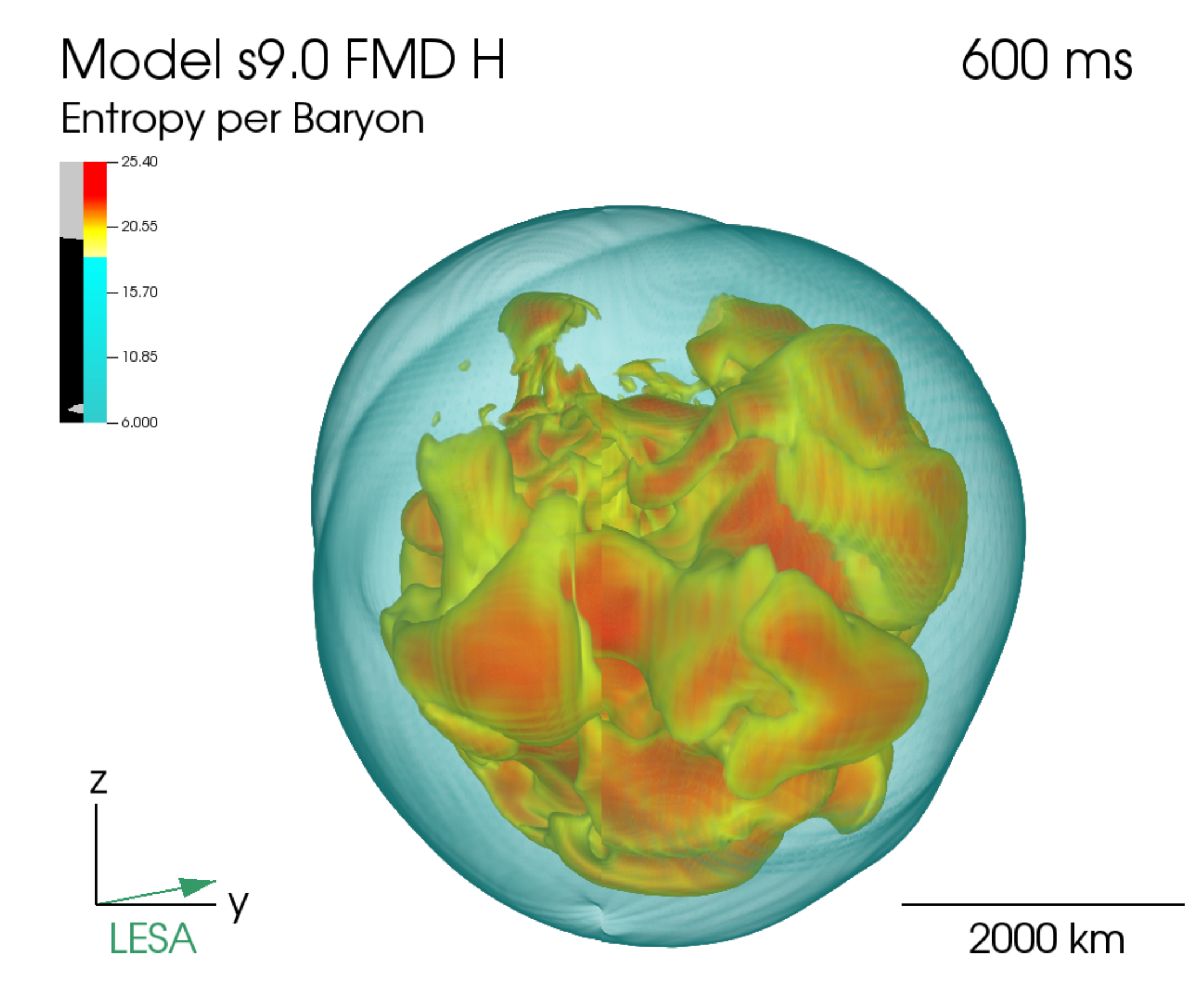}{0.48\textwidth}{}
          \fig{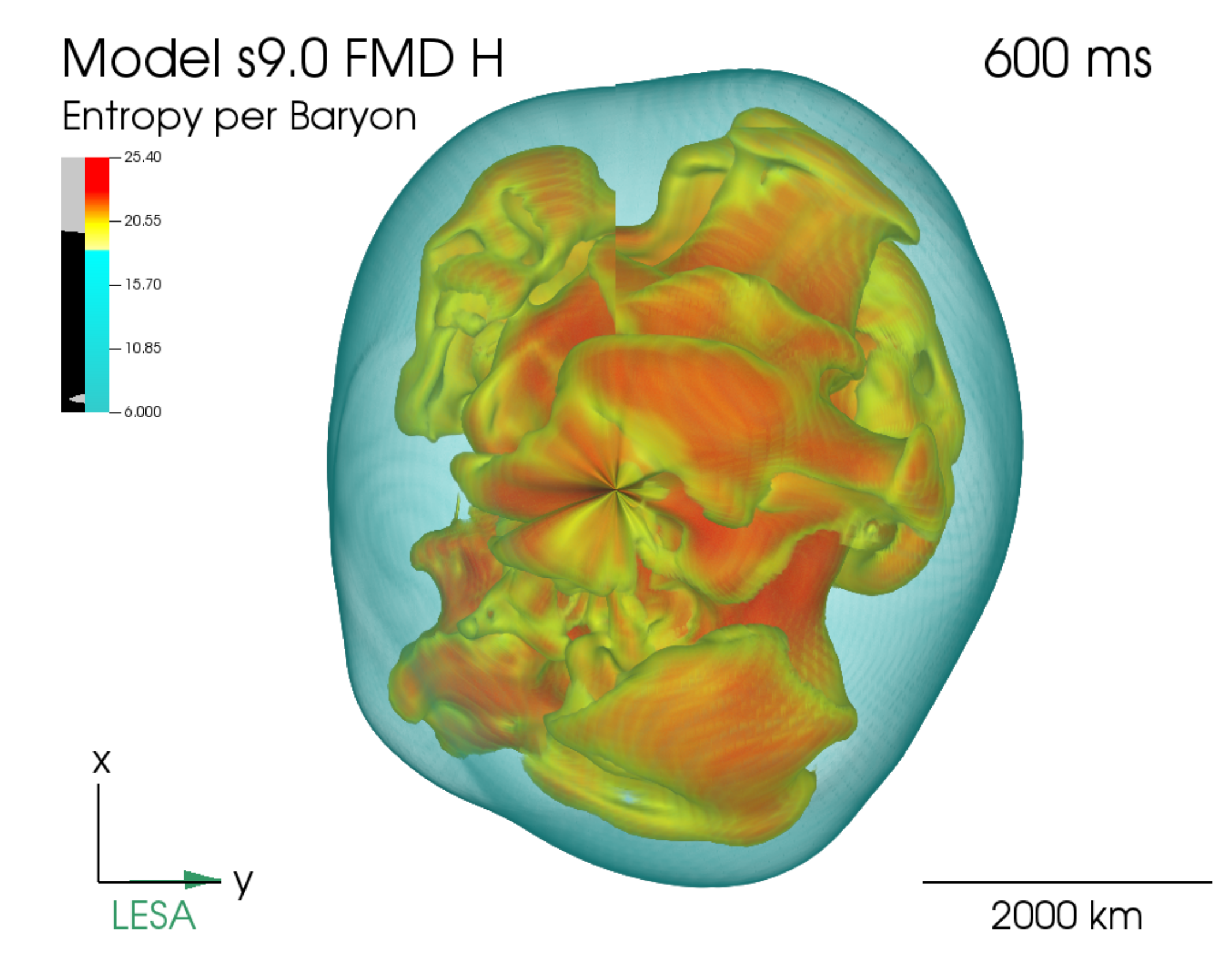}{0.48\textwidth}{}
         }
\caption{\new{Visualization of the explosion geometry by means of the entropy per baryon (in units of $k_{\mathrm{B}}$) for model `s9.0 FMD H'.
        The volume renderings show the shock front (turquoise layer) and plumes of neutrino-heated matter in the post-shock layer (yellow and orange surfaces)
        for a temporal snapshot at 600\,ms close to the end of the simulation.
        The left and right panels depict different planes (see the tripod in the lower left corner of each panel).
        We emphasize that the LESA direction nearly coincides with the $y$-axis (indicated by the green arrows in the tripods),
        and that there is no correlation between the LESA direction (see Figure~\ref{fig:lesa_volume_rendering_s20_s90})
        and the global deformation of the expanding post-shock material.
        The streak-like pattern in the center of the right panel is a well-known artifact for viewing directions along the polar axis,
        arising from the transformation from spherical polar coordinates to the Cartesian coordinate system of the visualization tool.}
        \label{fig:entropy_volume_rendering_s90}}
\end{figure*}

Overall, we conclude that the LESA phenomenon with its
characteristic features is shared by all of our 3D models.
This disproves speculations that LESA is a numerical artifact of the RbR+ 
approximation.
Our results are consistent with those reported by \citet{2014ApJ...792...96T},
\citet{2014PhRvD..90d5032T}, and \citet{2016ARNPS..66..341J},
where LESA was discussed as a lepton-number emission self-sustained asymmetry
first. Since a key feature of LESA is a dipole emission component that grows
relative to the monopole, it shall be noted that during the 
phases when the dipole dominates the higher-order multipoles,
the absolute value of the total electron lepton-number flux
(the monopole) in our s20 models is higher by a factor of ${\gtrsim\,2}$
than in the simulations with the \textsc{Vertex} code
presented by \citet{2014ApJ...792...96T}.
This discrepancy might stem from differences in the neutrino-interaction rates,
which are considerably more elaborate in many aspects in the calculations
with \textsc{Vertex}. It could also be a consequence of a different 
contraction behavior of the neutron star in response to different
energy and lepton-number loss through neutrino radiation.        
Moreover, proto-neutron star convection is substantially
stronger in calculations with the \textsc{Alcar} code as discussed by 
\citet{2018MNRAS.481.4786J}. All of these aspects can directly or
indirectly influence the production and the convective and radiative
transport of lepton number inside of the neutron star and can thus have
an effect on the relative strength of the monopole and dipole of the
electron lepton-number emission.

The detailed temporal evolution of LESA strongly depends on the simulation setup,
including the neutrino-transport method, the grid resolution, and the progenitor 
model. Although the onset of lepton-emission asymmetries is
seen at roughly the same time in all of our models, with $l \sim 4$--6 modes
showing the fastest growth, the subsequent evolution that leads to the emergence
of a dominant LESA dipole could depend also on stochastic initial seeds and a
stochastically-triggered complex feedback mechanism between asymmetric neutron-star
convection and mass accretion of lepton-rich material as described by 
\citet{2014ApJ...792...96T} and \citet{2016ARNPS..66..341J}. Further analysis
is needed, based on a future, larger pool of 3D simulations with varied
physics inputs and numerical treatments of hydrodynamics and neutrino transport.
In this context it is assuring that \citet{2018ApJ...865...81O} and
\citet{2019MNRAS.482..351V} also observed
the LESA phenomenon in their 3D simulation with two-moment transport including
velocity-dependent terms.

\section{LESA and Proto-Neutron Star Convection}
\label{sec:LESA-PNS}

A detailed description of the characteristic empirical features of the LESA
phenomenon was provided by \citet{2014ApJ...792...96T} and \citet{2016ARNPS..66..341J},
whose results were confirmed by the analysis of our current models in
Section~\ref{sec:lesa} \citep[see also][]{2018ApJ...865...81O,2019MNRAS.482..351V}.
As mentioned above and discussed in the previous works,
the major contribution to the dipole emission 
originates from the convective layer inside of the neutron star, whose 
hemispheric differences in the lepton-number loss rate can be further 
amplified by an asymmetry in the mass-accretion flow from the SN shock to the
neutron star.

\new{The directional dependence of the convective cell pattern and
of the lepton-number flux density in the interior of the neutron star can be seen
in the volume renderings of Figure~\ref{fig:lesa_volume_rendering_s20_s90}.
Both fiducial high-resolution models with FMD transport (`s20 FMD H', top panels, and `s9.0 FMD H', bottom panels)
show a pronounced hemispheric asymmetry for large (red) and small (blue) values of the lepton-number flux density
around the spherical region at roughly 10\,km (red sphere in the center),
measured at the times when the LESA dipole mode resides at the maximum value for each model.
The volume rendering was performed for the lepton-number flux density in the lab frame,
$F_{\mathrm{lnf}}^\mathrm{lab} $, which connects to the co-moving frame lepton-number flux density via Equations~\eqref{eq:lnf} and \eqref{eqn:flux_lab_frame}.
Since in the neutrino-diffusion regime (which well applies to the convective layer in the proto-neutron star)
the second term on the rhs of Equation~\eqref{eqn:flux_lab_frame} dominates by far,
the red bubbles in Figure~\ref{fig:lesa_volume_rendering_s20_s90} mainly visualize the velocities of the convective cells,
demonstrating a clear hemispheric asymmetry.
Because these asymmetries in the proto-neutron star are quite similar between the two fiducial simulations,
we pick only one of them (model `s20 FMD H') for the following analysis.}

\new{In the volume-rendered images of Figure~\ref{fig:lesa_volume_rendering_s20_s90},
the lepton-number flux density is displayed through semitransparent isosurfaces
that are picked by choosing non-vanishing transparency for selected flux-density values
(see the grey markers at the color bar of each panel).
The inner red sphere corresponds to the spherical lepton-number flux
that diffuses out from the non-convective central core.
The red plumes visualize the pattern of small-scale convective cells
that fill the convection layer of the proto-neutron star and
become visible because $F_\mathrm{lnf}^\mathrm{lab}$ depends on the radial fluid velocity
through the second term on the rhs of Equation~\eqref{eqn:flux_lab_frame}.
Stronger convection in the hemisphere pointing in the LESA direction
(which is indicated by the green arrows in the tripod of each panel)
and weaker convection in the opposite direction are obvious from
the larger red plumes in the former case and the smaller and less distinct plumes in the latter case.
These hemispheric differences in the strength of overturn activity in the proto-neutron star convection layer
lead to the pronounced dipolar asymmetry of the lepton-number flux exterior to the convective shell,
which is visualized by the enveloping, cap-like blue surfaces.}

\new{The bottom panels of Figure~\ref{fig:lesa_volume_rendering_s20_s90} together with Figure~\ref{fig:entropy_volume_rendering_s90} demonstrate
that the LESA direction has no obvious correlation with the deformation that develops in the expanding post-shock ejecta.
This can be understood by the fact that the LESA asymmetry is large (of order tens of percents) in the lepton-number flux,
but small (on the level of a few percent only) in the total number and energy flux of $\nu_e$ plus $\bar{\nu}_e$
(see \citealt{2014ApJ...792...96T,2014PhRvD..90d5032T}).
Therefore, since ($\nu_e$ plus $\bar{\nu}_e$)-emission asymmetries due to accretion have mostly a larger amplitude,
LESA has no dominant direct impact on the growth of neutrino-heated bubbles, and therefore on explosion asymmetries.}

\begin{figure*}
    \includegraphics[width=\textwidth]{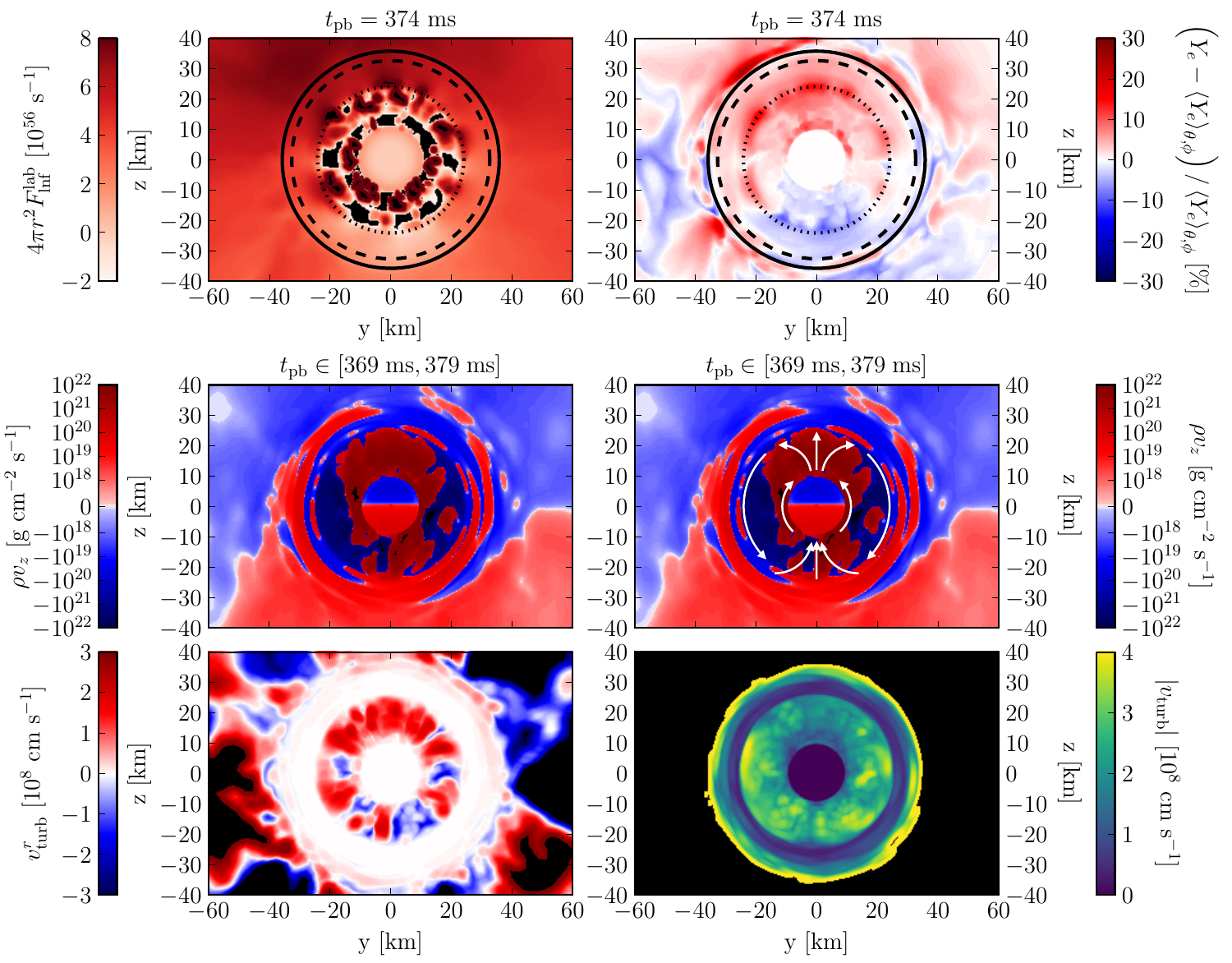}
    \caption{Asymmetries associated with the LESA phenomenon inside and around
            the neutron star in 3D model `s20 FMD H'. The six panels provide
            different quantities in the cross-sectional $y$-$z$-plane,
            which contains the polar axis of our spherical polar grid and cuts 
            through the neutron star very close to the axis of the LESA dipole
            (see Figure~\ref{fig:lesa_dipole_direction_aitoff_s20_s90}).
            \new{The LESA direction points (roughly) to the 12 o'clock position in all panels.}
            The upper left panel displays the radial component of the electon
            lepton-number flux $F_{\mathrm{lnf}}^{\mathrm{lab}}$,
            transformed to an observer in the rest frame of the stellar center
            (see Equation~\eqref{eq:lnf}) and scaled by the surface area
            $4\pi r^2$. The upper right panel shows spatial variations
            of the electron fraction $Y_{e}$, normalized by the angular 
            average $\langle Y_e(r)\rangle_{\theta,\phi}$.
            The white circular region around the center corresponds to the innermost 10\,km,
            which are computed in spherical symmetry.
            Both plots show a post-bounce time of
            ${t_{\mathrm{pb}} = 374\unitspace\mathrm{ms}}$.
            \new{Black} circles indicate the effective energy spheres of all
            neutrino species (as defined in Equation~(14) in Paper I),
            solid for \nue, dashed for \nubar, and dotted for \nux.
            The middle panels visualize the $z$-component of the mass-flux density, 
            $\rho v_{z}$, averaged over the time interval of
            ${t_{\mathrm{pb}} \in [369\unitspace\mathrm{ms}, 379\unitspace\mathrm{ms}]}$.
            The white arrows in the middle right panel sketch the underlying flow 
            pattern, forming a dipolar mode within the convective layer of
            the neutron star.
            \new{The lower panels show the cell structure in the convective layer
            by means of the turbulent radial velocity $v^r_{\mathrm{turb}}$ (left) and
            the absolute turbulent velocity $\left| v_{\mathrm{turb}} \right|$ (right),
            both averaged over the time interval of
            ${t_{\mathrm{pb}} \in [369\unitspace\mathrm{ms}, 379\unitspace\mathrm{ms}]}$.
            Values that exceed the range of the corresponding color bar of each panel appear in black.
            The global dipolar flow pattern visible in the middle panels,
            which encompasses the whole proto-neutron star convection layer,
            is superimposed on the familiar pattern of smaller-scale convective cells
            filling the convective shell as visible in the lower panels.
            These convective cells possess a hemispheric asymmetry in the LESA dipole direction
            with larger and more extended outflow structures on the side of the higher lepton-number flux,
            which is obvious from the bottom panels and which can also be concluded from the visualization
            of the lab-frame lepton-number flux density in Figure~\ref{fig:lesa_volume_rendering_s20_s90}
            and in the cross-sectional plots of Figure 10 in \citet{2014ApJ...792...96T}.}
            \label{fig:lesa_neutron_star_s20}}
\end{figure*}

Figure~\ref{fig:lesa_neutron_star_s20} displays a zoom of the
interior of the neutron-star and its immediate surroundings in our 3D model
`s20 FMD H'. The cross-sectional plane chosen for the plots is close to the
axis of the LESA dipole. 
The upper left panel shows the radial component of the lepton-number flux
in the laboratory frame (i.e., the rest frame of the stellar center),
$F_{\mathrm{lnf}}^{\mathrm{lab}}$ (see Equation~\eqref{eq:lnf}),
scaled with the surface area $4\pi r^2$, at $t_\mathrm{pb}=374$\,ms,
which is about the time when the LESA dipole reaches its maximum
(see Figure~\ref{fig:lesa_multipoles_cmap_s20_s90}, upper left panel).
The higher lepton-number flux in the northern 
hemisphere is correlated with an excess of the electron
fraction, $Y_e$, compared to its angular average at each radius 
(upper right panel).
This \new{pattern associated with LESA \citep{2014ApJ...792...96T}} grows within the convective
layer, which extends from the edge of the white circular region indicating
the 10\,km core that is computed in 1D, to the outer boundary of the
convective layer, which roughly coincides with the effective energy sphere
of muon and tau neutrinos (dotted \new{black} circle in the upper right panel). 
The LESA $\nu$-emission and $Y_e$ asymmetries continue within
the overlying radiative shell, where faster neutrino diffusion
carries the leptons to the neutrinospheres of $\nu_e$ and $\bar\nu_e$ 
(solid and dashed \new{black} circles, respectively), and persist into the 
accretion layer exterior to the neutrinospheres. 

The effective flow pattern that is responsible for these hemispheric 
differences in the lepton-number transport out from the deep inner core
of the proto-neutron star is displayed in the middle panels of 
Figure~\ref{fig:lesa_neutron_star_s20}, where the component of the
mass flux in the north-south ($z$) direction, $\rho v_z$,
is visualized by color coding.

\new{The bottom panels of Figure~\ref{fig:lesa_neutron_star_s20} show
the familiar cell pattern of the convective layer in the proto-neutron star.
The characteristic convective feature associated with LESA
is stronger outflow in the hemisphere pointing in LESA direction
and stronger convective inflow in the opposite hemisphere,
visible from the positive and negative radial components of the turbulent velocities
(defined by $v^r_{\mathrm{turb}} = v_r - \left< v_r \right>_{\theta, \phi}$,
where $\left< v_r \right>_{\theta, \phi}$ is the angular average of $v_r$)
in the bottom left panel.
In contrast, the absolute values of the convective velocities
($v_{\mathrm{turb}} = \sqrt{(v_r - \left< v_r \right>_{\theta, \phi})^2 + v_\theta^2 + v_\phi^2}$,
see bottom right panel of Figure~\ref{fig:lesa_neutron_star_s20})
may not display any clear correlation with the LESA dipole direction
(see also \citealt{2019arXiv190106235W}).}

Superimposed on the smaller-scale convective
variations (see the pattern of convective cells in the upper left panel
\new{and in the bottom panels of Figure~\ref{fig:lesa_neutron_star_s20}}),
a global dipolar flow can be identified, indicated by
the white arrows in the middle right panel: Around the north pole 
lepton-rich plasma rises through the convective layer, transporting
electron-lepton number from the inner core to regions closer to the
neutrinospheres, where the high electron abundance causes enhanced 
$\nu_e$ emission. Matter with decreasing $Y_e$ streams along the outer
edge of the convective layer from the north pole towards the south pole, 
further losing lepton number by radiating $\nu_e$. Near the south
pole the flow of neutron-rich, specifically (i.e., per nucleon) heavier and thus denser matter submerges
again to deeper layers, where it is channelled into
currents that stream back towards the northern hemisphere along the 
inner border of the convective layer. On its way from south to north,
the plasma is
refuelled with fresh leptons by absorbing neutrinos diffusing out from
the high-density core.

A deficiency of the present models in connection to this analysis of 
LESA should be mentioned here. The 1D core with a chosen radius of
10\,km, which is employed in our 2D and 3D simulations
for reasons of computational efficiency,
turned out to be too large during the later post-bounce evolution of
our models. At times beyond about 200\,ms after bounce, the inner
boundary of the convective layer hits the radius of the 1D core
(see upper left panel of Figure~\ref{fig:neutron_star_rayleigh_S20_3D}),
and correspondingly, the spherically symmetric treatment of the 1D
core prevents the convective shell to grow deeper.
The 2D calculations of \citet{2018MNRAS.481.4786J}, where the s20
progenitor was run with the same input physics but
with a considerably smaller 1D core,
demonstrate that the inner boundary of the
convective shell reaches a radius of
$\sim$10\,km already after about 150\,ms and becomes as small as
$\sim$7\,km after 500\,ms.
\new{We will confirm this finding in a similar analysis of two 2D simulations in the following section.}
It is therefore clear that the use of the
spherical core has constrained the later growth of the 
proto-neutron star convection layer in the simulations of the present
paper. Potentially, this might even have prevented a growth of the
LESA dipole beyond the amplitudes obtained in our models.
Nevertheless, the basic features of the LESA phenomenon
agree well with the results reported by \citet{2014ApJ...792...96T}
and \citet{2016ARNPS..66..341J} from calculations with the
\textsc{Prometheus-Vertex} code,
where the 1D core was as small as $\sim$1.5\,km and had no impact 
on the development of neutron-star convection. Evaluating the 
available set of 3D models from the \textsc{Prometheus-Vertex} runs,
\citet{2015StockingerMaster} also observed a dipolar 
flow pattern that encompasses the convective layer in the neutron
star and transports electron lepton-number from one hemisphere to
the other, very similar to the one visible in 
Figure~\ref{fig:lesa_neutron_star_s20}. 

\section{Physics Origin of the LESA Phenomenon}\label{sec:lesa_explanation}
\label{sec:LESA-Chandra}

What is the physical mechanism that leads to the LESA
lepton-emission asymmetry? Why does the dipole mode become the
dominant multipole component after a few 100\,ms of post-bounce 
evolution? In the following we intend to further illuminate the 
development of convection in the proto-neutron star, where the major
contribution to LESA is produced. For this purpose
\new{we explore, tentatively and experimentally, whether there could be an analogy between the occurrence of different spherical harmonics modes
and the possibility of a dominant dipole in the convective proto-neutron star shell of our models on the one hand,
and the excitation of harmonics of different orders in thermally unstable spherical shells as theoretically discussed by \citet{1961hhs..book.....C} on the other hand.
Since the LESA lepton-number emission dipole asymmetry mainly originates from the proto-neutron star convection zone,
we consider such a connection as a not too far-fetched possibility.
Although such a link} cannot provide a satisfactory explanation of the
properties of the system in the fully nonlinear stage, we will be able to
demonstrate that the different $l$-components in the mode spectrograms of
the lepton-number flux (Figure~\ref{fig:lesa_multipoles_cmap_s20_s90})
exhibit a growth behavior compatible with basic expectations from
Chandrasekhar's linear theory. Besides the fact that this connection
may be considered as a check
of physical consistency, our analysis also allows for speculations 
about the subsequent evolution of the system in view of the trends 
displayed by its characteristic parameters.

Building up on the seminal work by Lord Rayleigh, \citet{1961hhs..book.....C}
determined the critical conditions for the onset of thermal instability in fluid 
spheres and spherical shells by linear stability analysis of a variety of 
situations, invoking simplifying assumptions (e.g., the 
Boussinesq approximation, simple radial dependences of gravity and temperature,
and constant medium-dependent coefficients such as those for volume 
expansion, viscosity, specific heat, or thermal conductivity).
He constitutes that, provided the principle of the exchange of stabilities is
valid (i.e., that the transition from stability to instability occurs via a
stationary pattern of fluid motions instead of oscillatory motions), instability 
will set in for disturbances of the spherical harmonics mode $l$ with
the lowest value of the critical Rayleigh number. This means that at
marginal stability the convective pattern will manifest itself in this
$l$-mode when the Rayleigh number ${\cal R}$ of the system, 
\begin{equation}\label{eqn:rayleigh_number}
{\cal R} = \alpha\, \left| \frac{\mathrm{d}\Phi_\mathrm{g}}{\mathrm{d}r}\right|
\,\left|\frac{\mathrm{d}T}{\mathrm{d}r}\right|
\left( \xi\, \nu_{\mathrm{N}} \right)^{-1}
\left( r_{\mathrm{o}} - r_{\mathrm{i}} \right)^{4},
\end{equation}
exceeds the critical Rayleigh number ${\cal R}_{{\mathrm C},l}$ of the mode.
In Equation~\eqref{eqn:rayleigh_number}, $\Phi_{\mathrm{g}}$ is the gravitational
potential, $T$ the temperature, $\alpha$ the coefficient of volume expansion,
$\xi$ the thermometric conductivity, and $\nu_\mathrm{N}$ the kinematic 
viscosity. Since in our case of application viscosity is dominated by numerical
effects, we use the subscript ``N'' for the kinematic viscosity.
The radii $r_\mathrm{i}$ and $r_\mathrm{o}$ are the inner and outer boundaries 
of the convective layer, respectively.

With this definition of the Rayleigh number the critical condition means
that instability sets in at the minimum temperature gradient at which a balance
can be maintained between the kinematic energy dissipation by viscosity and
the internal energy liberated by the buoyancy force \citep{1961hhs..book.....C}.
It further means that for Rayleigh numbers lower than the critical 
Rayleigh number ${\cal R}_{{\mathrm C},l}$ of a mode $l$, disturbances of this
mode are expected to be stable, while such disturbances will become unstable
when the Rayleigh number exceeds ${\cal R}_{{\mathrm C},l}$.

We attempt now to verify these aspects from our numerical results,
\new{despite being aware of a number of severe caveats.
In particular, we draw an analogy between thermally driven convection as discussed by Chandrasekhar
and Ledoux convection as relevant for the proto-neutron star environment.
Moreover, we apply Chandrasekhar's arguments to the phases of beginning convection
as well as fully developed convective activity in our 3D models, which is likely to overstretch the applicability of Chandrasekhar's linear theory.
Therefore, the arguments presented in the following are unquestionably very tentative and experimental.}

\begin{figure*}
    \includegraphics[width=\textwidth]{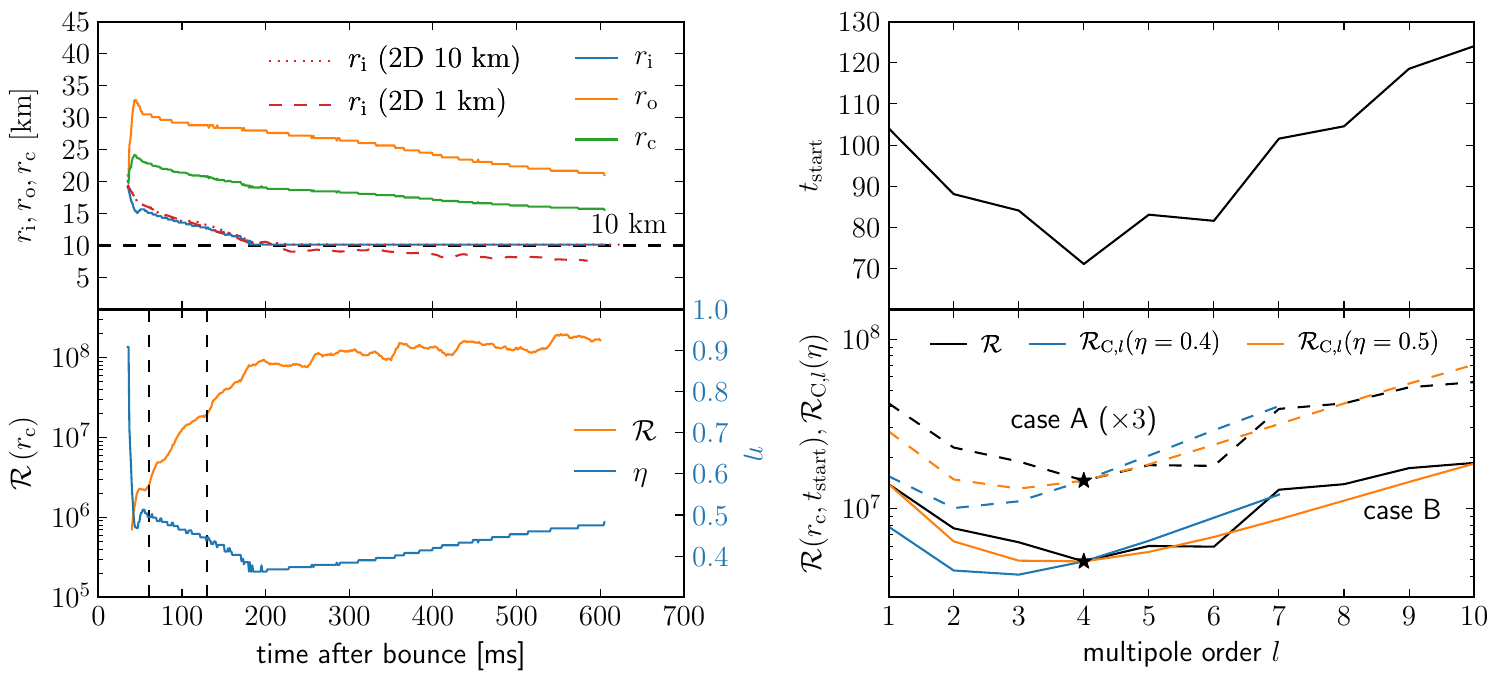}
    \caption{Analysis of the convective layer in the proto-neutron star of 3D model `s20 FMD H'
            in view of the linear discussion by \cite{1961hhs..book.....C} of the onset
            of convective (thermal) instability in spherical shells.
            Upper left panel: Inner, outer, and central radius of the convective region,
            $r_\mathrm{i}$ (blue), $r_\mathrm{o}$ (orange), and $r_\mathrm{c} = 
            0.5(r_\mathrm{i}+r_\mathrm{o})$ (green), respectively, as functions of post-bounce time.
            The horizontal dashed line at 10\,km indicates the boundary of the core that is
            computed in spherical symmetry instead of 3D for reasons of numerical efficiency.
            \new{To demonstrate the impact of the inner boundary,
            we plot $r_\mathrm{i}$ also for two 2D reference simulations with
            spherical symmetric cores of 10\,km (red dotted line) and 1\,km (red dashed line), respectively.}
            Lower left panel: Rayleigh number $\mathcal{R}(r_\mathrm{c})$, 
            evaluated at the central radius 
            $r_{\mathrm{c}}$ and smoothed by running averages over ${10\unitspace\mathrm{ms}}$
            (orange), and radius ratio $\eta = r_\mathrm{i}/r_\mathrm{o}$ (blue) as
            functions of post-bounce time. The vertical dashed lines indicate the boundaries of
            the time interval ($60\unitspace\mathrm{ms}$ to $130\unitspace\mathrm{ms}$) for 
            the initial growth of multipole modes of the lepton-number flux 
            between $l=1$ and $l=10$ as shown in the right panel.
            Upper right panel: Post-bounce times $t_\mathrm{start}(l)$ vs.\ multipole order $l$
            with $t_\mathrm{start}(l)$ representing the instants when the multipole
            coefficients of the electron lepton-number flux, $A_{\mathrm{lnf}}$ (see
            Figure~\ref{fig:lesa_multipoles_cmap_s20_s90}), exceed a threshold value of 
            $5.5 \times 10^{54}\unitspace\mathrm{s}^{-1}$ for the first time.
            Lower right panel: Rayleigh numbers ${\cal R}(r_\mathrm{c},t_\mathrm{start})$ (black
            lines) evaluated at $r_\mathrm{c}$ (lower left panel) at the onset times 
            $t_{\mathrm{start}}$ (shown in the upper right panel) of the different multipoles 
            $l$ of the electron lepton-number flux, compared to Chandrasekhar's results 
            of the critical Rayleigh numbers $\mathcal{R}_{\mathrm{C},l}(\eta)$ for values of
            $\eta = 0.4$ (blue lines) and $\eta = 0.5$ (orange lines). 
            The solid lines of case~B correspond to Chandrasekhar's case for rigid
            inner and free outer boundary, the dashed lines of case~A to Chandrasekhar's case with
            free inner and outer boundaries (shifted by a factor of 3 for better visibility).
            To adapt to our problem, we have scaled Chandrasekhar's critical Rayleigh numbers
            such that the value for $l=4$ coincides with the minimum of 
            ${\cal R}(r_\mathrm{c},t_\mathrm{start})$ that we obtained at this mode number 
            (black star).
            \label{fig:neutron_star_rayleigh_S20_3D}}
\end{figure*}

In Figure~\ref{fig:neutron_star_rayleigh_S20_3D}, left panels, we provide
information on the evolution of the outer boundary, $r_\mathrm{o}$,
and inner boundary, $r_\mathrm{i}$, of the
neutron-star convection layer in our 3D model `s20 FMD H', 
on the central radius of this layer,
$r_\mathrm{c} = 0.5(r_\mathrm{o} + r_\mathrm{i})$, and on the
radius ratio $\eta = r_\mathrm{i}/r_\mathrm{o}$. Here, $r_\mathrm{i}$
is defined as the radius at which the angle-averaged, non-radial velocities
exceed $10^{7}\unitspace\mathrm{cm}\unitspace\mathrm{s}^{-1}$, whereas
$r_\mathrm{o}$ is determined as the radial position where the angle-averaged, 
non-radial velocity drops to roughly 40\% of its maximum value in the
convective shell. The latter condition allows for an unambiguous identification
of the outer shell boundary also at times when accretion downflows create
substantial velocity perturbations in the convectively stable layer that
separates the convective shell inside of the neutron star from the turbulent
flows in its surroundings.

\new{We emphasize that the use of a 1D core as mentioned in Section 4 has no influence on the analysis discussed here,
whose results are shown in Figure~\ref{fig:neutron_star_rayleigh_S20_3D},
because during the time interval of this analysis the inner radius of the proto-neutron star convection layer is still well outside of the 1D core.
To assess the impact of the spherical core on the inner boundary of the convection layer at later times,
we show $r_\mathrm{i}$ in the upper left panel of Figure~\ref{fig:neutron_star_rayleigh_S20_3D}
also for two 2D reference simulations of the s20 progenitor model,
one with a spherically symmetric core of the same size of 10\,km as in the 3D simulations (red dotted line),
and the other one with a much smaller core of only 1\,km (red dashed line).
In the latter case, the region of convection moves only very slowly into the neutron star down to 7--7.5\,km
by the end of the simulation at 600\,ms after bounce.
In contrast, the 2D simulation with a 10\,km core behaves similarly to the 3D simulation,
with $r_\mathrm{i}$ touching 10\,km only close to 200\,ms.
The analysis following below, however, focuses on the development of the proto-neutron star convection in the first 60 to 130\,ms after bounce,
and should therefore not be influenced by the constraints arising from the spherical core.
Later on, a slightly more extended convective layer, i.e.\ somewhat larger values of $r_\mathrm{o} - r_\mathrm{i}$,
will even strengthen the conclusions that we will come up with.}

The lower left panel of Figure~\ref{fig:neutron_star_rayleigh_S20_3D}
also shows the Rayleigh number ${\cal R}(r_\mathrm{c})$, evaluated at 
$r_\mathrm{c}$, as a function of post-bounce time. ${\cal R}$ is calculated 
according to Equation~\eqref{eqn:rayleigh_number}, using angular averages
for all quantities needed. The coefficient of volume expansion, $\alpha$,
can be derived from the equation of state, 
\begin{equation}
\alpha = \left|\frac{1}{\rho}\left(\frac{\partial\rho}{\partial T}\right)_{\! P}\right|
= \left|-\,\frac{1}{\rho}\left(\frac{\partial P}{\partial T}\right)_{\! \rho}
\left(\frac{\partial P}{\partial \rho}\right)_{\! T}^{\! -1}\right| \,,
\label{eq:alpha}
\end{equation}
with $P$, $\rho$, and $T$ being pressure, density, and temperature. For the
thermometric conductivity $\xi$ we use, 
as a proxy\footnote{The heat flux $\vect{F}_\mathrm{heat} = k\vect{\nabla}T$,
with $k$ being the coefficient of heat conduction, can be identified with the 
total neutrino flux density (accounting suitably for all neutrino species),
$\vect{F}= -D\vect{\nabla}E = -D(\partial E/\partial T)_\mu\vect{\nabla}T$, 
where
$D$ is the diffusion coefficient and $\mu$ the neutrino chemical potential.
The latter relation holds under the assumption of equilibrium diffusion, in
which case $T_\nu = T$, and for negligible chemical potential gradients.
Therefore $k = D(\partial E/\partial T)_\mu$. Using the definition
$\xi = k(\rho c_V)^{-1}$ \citep{1961hhs..book.....C} and 
$\rho c_V = (\partial E_\mathrm{int}/\partial T)_\rho$ with $c_V$ being the 
specific heat (per unit of mass) at constant volume and $E_\mathrm{int}$ the
internal energy density including the contributions from stellar plasma 
and neutrinos, one gets: $\xi = D\,(\partial E/\partial T)_\mu
(\partial E_\mathrm{int}/\partial T)_\rho^{-1}$.
We assume that the factor $(\partial E/\partial T)_\mu 
(\partial E_\mathrm{int}/\partial T)_\rho^{-1}$ is roughly a constant,
and thus we can identify $\xi \sim D$, which, for the radial component,
yields $\xi \sim D = \left| F^r\,(\partial E/\partial r)^{-1}\right|$.},
an effective diffusion coefficient
for the total energy flux carried by neutrinos (summed over all
neutrino species $\nu$):
\begin{equation}\label{eqn:thermal_diffusivity_neutrinos}
\xi = \left| \left( \sum_{\nu} \int F^{r}_{\nu} \dint \varepsilon \right)
\left(\frac{\partial}{\partial r} \left[ \sum_{\nu} \int E_{\nu} 
\dint \varepsilon \right] \right)^{-1}\right|\,,
\end{equation}
where $E_\nu$ and $F_\nu^r$ are measured in the co-moving frame of the 
stellar plasma. In the kinematic viscosity $\nu_\mathrm{N}$ appearing in
Equation~\eqref{eqn:rayleigh_number}, we consider the contributions
from neutrino viscosity as well as numerical viscosity associated with the
hydrodynamics solver. The latter is usually dominant by a factor of 
$\sim$20--100, depending on the time and position in the convective shell.
The neutrino viscosity, $\nu_{\nu}$, is evaluated by following 
\citet{1996ApJ...473L.111K} and \citet[see Equation~(8) there]{2015MNRAS.447.3992G}
and is found to range between some $10^9$\,cm$^2$\,s$^{-1}$ and 
nearly $10^{11}$\,cm$^2$\,s$^{-1}$, in agreement with estimates in the
mentioned references. In contrast, the numerical Reynolds number during
the relevant time is around 200--250 (employing Equation~(6.36) of
\citealt{2016MelsonPhD}, in rough agreement with crude estimates by
\citealt{1996ApJ...473L.111K}). Using 
\begin{equation}\label{eqn:reynolds_number}
\mathrm{Re} = \frac{\Delta r\,\, v_\mathrm{turb}}{\nu_{\mathrm{num}}}
\end{equation}
with $\Delta r = r_\mathrm{o} - r_\mathrm{i}\sim 15$\,km and
maximal turbulent velocities in the convective layer of 
$v_\mathrm{turb} = \sqrt{(v_r - \langle v_r\rangle_{\theta,\phi})^2 +
v_\theta^2 + v_\phi^2} \sim 1.4\times 10^8$\,cm\,s$^{-1}$,
we obtain $\nu_\mathrm{num} \sim 10^{12}$\,cm$^2$\,s$^{-1}$.
Therefore we get $\nu_\mathrm{N} = \nu_\nu + \nu_\mathrm{num} \sim
\nu_\mathrm{num}$ within the convective layer of the proto-neutron
star.
\new{Considering again Equation~\eqref{eqn:rayleigh_number},
we see that the overestimation of the kinematic viscosity associated with the numerical viscosity in our simulations
leads to an effective decrease of the Rayleigh number by roughly two orders of magnitude.
For more realistic values of the kinematic viscosity associated with neutrinos, we expect even larger Rayleigh numbers and,
probably, proto-neutron star convection to occur even more readily.
In the phase of active convection, a lower viscosity will have an effect on the power spectrum of turbulent kinetic energy,
in particular it will permit a larger inertial range between the energy-injection scale and the viscous dissipation scale.
This, however, should not play a major role for the discussion of LESA, because LESA is mainly a large-scale phenomenon
(with dominant dipole and quadrupole modes of asymmetry),
whereas the smallest resolved scales (of a few grid cells) in the simulations are unlikely to add any relevant physics for LESA.
Therefore, we conclude that the dominance of the numerical viscosity in our 3D models
is an acceptable compromise between feasibility and physical accuracy,
which cannot be overcome before significantly better resolved 3D models become affordable.
This compromise should not change the basic physics discussed here.}

The Rayleigh number ${\cal R}(r_\mathrm{c})$ 
(Figure~\ref{fig:neutron_star_rayleigh_S20_3D}, lower left panel)
exhibits a steep, monotonic rise
during the time interval between $\sim$60\,ms and $\sim$130\,ms 
after bounce, when convection in the proto-neutron star of model
`s20 FMD H' gains strength and asymmetry modes of the electron
lepton-number flux with multipole orders from
$l = 1$ to $l = 10$ develop significant amplitudes for the first time
(see Figure~\ref{fig:lesa_multipoles_cmap_s20_s90}).
For defining the onset times $t_\mathrm{start}(l)$ we choose 
a threshold value for the multipole coefficients $A_\mathrm{lnf}$
of $5.5 \times 10^{54}\unitspace\mathrm{s}^{-1}$,
which corresponds to the level where the initial multipole structures can be identified in Figure~\ref{fig:lesa_multipoles_cmap_s20_s90}.
The start times as function of $l$
are given in the upper right panel of 
Figure~\ref{fig:neutron_star_rayleigh_S20_3D}. Now picking the
Rayleigh numbers ${\cal R}(r_\mathrm{c})$ at times $t_\mathrm{start}(l)$
from the lower left panel of Figure~\ref{fig:neutron_star_rayleigh_S20_3D},
we can plot the black (solid and dashed) lines in the lower right panel
of this figure. 

These values of ${\cal R}(r_\mathrm{c},t_\mathrm{start}(l))$ \new{can be compared to}
the critical Rayleigh numbers ${\cal R}_{\mathrm{C},l}$ derived
by \citet{1961hhs..book.....C} for the two shell-radius ratios
of $\eta = 0.4$ and $\eta = 0.5$, which are in the
ballpark of our model results (see the blue line in the lower 
left panel of Figure~\ref{fig:neutron_star_rayleigh_S20_3D}).
We refer the reader to \citet[][\new{Tables~XX, XXII, and XXIII, as well as Figure~58}]{1961hhs..book.....C} for explicit
expressions and plotted as well as tabulated values.
We are only interested in
comparing the shapes of the curves ${\cal R}(r_\mathrm{c})$ and
${\cal R}_{\mathrm{C},l}$ as functions of mode-number $l$,
because Chandrasekhar's linear analysis of thermal instability
considered largely different physical conditions adequate for
thermal instability in laboratory
and terrestrial environments, and because our evaluation of 
${\cal R}(r_\mathrm{c})$ disregards potentially relevant numerical
factors (e.g., when using $\xi \sim D$). Therefore we 
scale Chandrasekhar's critical Rayleigh numbers such that the
value for $l=4$ coincides with the minimum obtained for our
Rayleigh numbers ${\cal R}(r_\mathrm{c})$ at this multipole
mode, indicated by black asterisks in the lower right panel
of Figure~\ref{fig:neutron_star_rayleigh_S20_3D}.

We consider two of the four cases of boundary conditions
employed by \citet{1961hhs..book.....C}, namely on the one
hand free surfaces (where tangential viscous stresses are
assumed to vanish) at $r_\mathrm{i}$ and $r_\mathrm{o}$
(case~A displayed by dashed colored lines in 
Figure~\ref{fig:neutron_star_rayleigh_S20_3D} and shifted by
a factor of 3 for better visibility) and on the other hand
a rigid surface at $r_\mathrm{i}$ (where the nonradial
components of the velocity vanish) and a free surface at 
$r_\mathrm{o}$ (case~B, shown with solid colored lines in 
the lower right panel
of Figure~\ref{fig:neutron_star_rayleigh_S20_3D}). Although
not motivated by the physical conditions in the proto-neutron
star, we also picked the second case because the 1D core used
in our simulations has a non-negligible influence on the
development of the convective layer by damping or even 
suppressing nonradial fluid motions close to the 3D/1D boundary.

The lower right panel of Figure~\ref{fig:neutron_star_rayleigh_S20_3D} 
thus allows us to compare our Rayleigh numbers
${\cal R}(r_\mathrm{c},t_\mathrm{start}(l))$ at the times when
the different multipole modes of the lepton-number flux 
between $l=1$ and $l=10$ begin to grow (black lines) with the 
(rescaled) critical Rayleigh numbers ${\cal R}_{\mathrm{C},l}$ when
disturbances of such $l$-modes are expected to become convectively unstable
(blue and orange lines for $\eta = 0.4$ and $\eta = 0.5$, respectively)
according to the linear analysis by \citet{1961hhs..book.....C}.
Interestingly, the shapes of the corresponding curves
exhibit good overall resemblance.
In particular case~B for $\eta = 0.5$ follows the 
trend of ${\cal R}(r_\mathrm{c},t_\mathrm{start}(l))$ \new{quite}
well, with a minimum in both cases being located at $l=4$, 
where a substantial value of $A_\mathrm{lnf}$ can be observed
first in Figure~\ref{fig:lesa_multipoles_cmap_s20_s90}. The value 
of $\eta = 0.5$ is indeed very close to the boundary-radius ratio for
the convective layer in time interval
$60\,\mathrm{ms}\lesssim t_\mathrm{pb}\lesssim 130\,\mathrm{ms}$, 
and the better match for case~B makes perfect
sense in view of our numerical constraints at the inner boundary
radius $r_\mathrm{i}$.

The location of the
minimum of the critical Rayleigh line depending on the shell
thickness, i.e.\ the value of $\eta$, can be roughly understood
by a simple argument. For a shell thickness
$\Delta r = (1-\eta)\eta^{-1}r_\mathrm{i}$ the preferred wavelength 
of a convective cell is $2\Delta r$. With $r_\mathrm{i} =
2\eta(1+\eta)^{-1}r_\mathrm{c}$ one obtains a preferred wavelength
of $\lambda_\mathrm{pref}(\eta) = 4(1-\eta)(1+\eta)^{-1}r_\mathrm{c}$, 
which yields $\lambda_\mathrm{pref}(0.4) = (12/7)r_\mathrm{c}$ and
$\lambda_\mathrm{pref}(0.5) = (4/3)r_\mathrm{c}$. On the other hand,
the wavelength of an $l$-mode ($m=0$) on a circle of circumference
$2\pi r_\mathrm{c}$ corresponds to $\lambda_l = 2\pi r_\mathrm{c} l^{-1}$.
Requesting $\lambda_\mathrm{pref}(\eta) \sim \lambda_l$, one finds that 
$l = 3$ fits better for $\eta = 0.4$ and $l=4$ matches better 
$\eta = 0.5$.

These results therefore suggest that the $l$-mode pattern in the 
electron lepton-number emission, emerging in model `s20 FMD H' around
100\,ms p.b.\ (Figure~\ref{fig:lesa_multipoles_cmap_s20_s90}),
is compatible with Chandrasekhar's linear theory for the onset of 
thermal convection in spherical shells applied to the convective
layer in the neutron star. While initially convective modes around
$l\sim 4$ are favored in their growth, because their critical
Rayleigh numbers are lowest, the monotonic rise of the Rayleigh
number within the contracting neutron star also permits dipolar
and quadrupolar modes to appear after $\sim$100\,ms. Since the
Rayleigh number increases further by about an order of magnitude, 
the later dominance of the $l=1$ convective mode
(Figure~\ref{fig:lesa_neutron_star_s20}) and of the
corresponding LESA dipole (Figures~\ref{fig:lesa_flux_aitoff_s20_s90}
and \ref{fig:lesa_multipoles_cmap_s20_s90}) is certainly not in 
conflict with the critical behavior of thermally unstable shells
as discussed by \citet{1961hhs..book.....C}.

However, Chandrasekhar's theoretical assessment does not yield
any prediction of the evolution of the convection layer in the
nonlinear phase and, moreover, it does not provide any growth rates
for the different convective $l$-modes. Therefore, it only allows
one to conclude that the conditions at $t\gtrsim 100$\,ms post bounce
are favorable also for the presence of the $l=1$ mode, but it does not
give an argument why the $l=1$ mode indeed becomes the dominant 
convective and lepton-emission asymmetry during long phases of the 
evolution.
In fact, follow-up work points to the need of a nonlinear treatment 
to answer the question of the pattern of convection in spherical
shells \citep[e.g.,][]{1975JFM....72...67B}.

Our connection to Chandrasekhar's analysis can only be considered
as tentative, because we invoked a number of shortcuts. Convection
inside nascent neutron stars is driven by entropy and lepton-fraction
gradients, captured by the Ledoux criterion, and our evaluation in
Section~\ref{sec:LESA-PNS} highlights the importance of variations
of the lepton fraction in the stellar plasma and in the neutrino
diffusion fluxes. A rigorous analytic assessment must take this
into account and must thus generalize the mathematical treatment
by \citet{1961hhs..book.....C}, which was focussed on the problem
of thermal instability driven by temperature gradients and the
associated heat conduction. The
critical $l$-mode-dependent Rayleigh numbers for the more complex, 
more general situation of proto-neutron star convection still have
to be derived.

An interesting implication of our discussion is the flashlight on
the role of numerical viscosity. With the resolution used in our
simulations, it is clearly dominant over the neutrino viscosity by
a significant factor (one to two orders of magnitude).
Therefore the numerical viscosity determined our estimated
values of the Rayleigh numbers for neutron-star conditions. This
suggests that higher values of the numerical viscosity might
artificially damp or suppress the appearance of dipolar and
quadrupolar flows, because the onset of these modes is 
constrained by higher values of the critical Rayleigh number.
This fact should be kept in mind when LESA is searched for in
the results from numerical calculations.

Nonlinear effects in the context of LESA beyond Chandrasekhar's
mathematical theory were described by \citet{2014ApJ...792...96T}
and \citet{2016ARNPS..66..341J}.
In order to explain the development of a dominant dipolar mode in
neutron-star convection and lepton-emission, and the long-time
stability of this phenomenon, 
\citet{2014ApJ...792...96T} proposed a coupling to a
hemispheric difference of the accretion flow between
shock and neutron star. Because of the harder spectra of $\bar\nu_e$
compared to $\nu_e$, the LESA dipole leads to stronger neutrino
heating in the direction of the higher $\bar\nu_e$ flux, i.e.,
opposite to the LESA dipole direction. The stronger heating pushes
the SN shock to larger radii and channels the mass-accretion flow
preferentially to the opposite hemisphere. The higher accretion
on this side of the neutron star brings fresh leptons mainly
to the hemisphere that is already lepton-rich,
and thus it strengthens the LESA dipole emerging from the
neutron-star convection layer. 

Because LESA is also found in neutron stars from low-mass progenitors,
where post-bounce accretion plays a relatively unimportant role, 
\citet{2016ARNPS..66..341J} suggested 
another feedback mechanism that could be responsible for the 
steep growth of the LESA dipole. These authors discussed
an amplification effect caused by a changing sign of the
thermodynamic derivative $(\partial\rho/\partial Y_\mathrm{lep})_{P,s}$
(with $s$ being the entropy per baryon) that appears in association with
the gradient of the lepton fraction in the Ledoux criterion for
convective instability. The sign 
changes from negative to positive when the medium deleptonizes,
with the threshold value of $Y_\mathrm{lep}$ being higher 
for higher density and temperature. When $Y_\mathrm{lep}$ drops
below the critical value,
a negative lepton-fraction gradient becomes stabilizing
instead of destabilizing, thus damping convective activity instead
of driving it (because only a negative entropy gradient remains 
as a driving term for the Ledoux convection). 
As a consequence, the convective flow is attenuated
especially in the hemisphere opposite to the LESA direction, where the
electron fraction and the total lepton fraction have dropped to lower 
values (see lower left panel in Figure~\ref{fig:lesa_neutron_star_s20}).
This weakens the convective transport of lepton number out from the 
diffusive inner core in this hemisphere, thus enhancing
the LESA dipole asymmetry. Such a positive feedback cycle could
gradually amplify the asymmetry and therefore might
explain why at some point in the evolution the dipole amplitude of
LESA increases monotonically. Since the electron fraction of the stellar
plasma as a function of density and temperature plays a role
(through its influence on the sign of 
$(\partial\rho/\partial Y_\mathrm{lep})_{P,s}$), this picture might
also explain why differences in the neutrino treatment
(e.g., FMD vs.\ RbR+ or different neutrino opacities) as well as
the nuclear equation of state (which determines the thermodynamic
derivative) can have an impact on the growth of LESA 
\citep[see][for a report on equation-of-state dependent 
differences]{2016ARNPS..66..341J}.

\section{Conclusions}
\label{sec:conclusion}

We presented an in-depth analysis of the effects of LESA in a set of eight 3D SN runs for two progenitors published in Paper~I.
LESA was first witnessed as a dipolar electron lepton-number emission asymmetry by \citet{2014ApJ...792...96T}
in their 3D simulations with ray-by-ray-plus (RbR+) neutrino transport, which neglects the nonradial components of the neutrino fluxes.
\new{Recently, LESA was also seen in one 3D model with fully multi-dimensional (FMD) neutrino transport by \citet{2018ApJ...865...81O}
and, possibly, by \citet{2019MNRAS.482..351V}.}
We observed the appearance of a strong LESA dipole mode after some 100\,ms p.b.\ in all of our models,
for exploding 9\,$M_\odot$ and nonexploding 20\,$M_\odot$ cases, both
with lower and doubled grid resolution, and with FMD as well as RbR+ neutrino transport.

All characteristic features of LESA reported by \citet{2014ApJ...792...96T}
and \citet{2014PhRvD..90d5032T}
were also diagnosed in the present models: (1) phases where the dipole
component of the lepton-number flux clearly dominates all higher-order 
multipoles and can reach even the strength of the monopole; 
(2) a relatively stable position or only slow
migration of the dipole direction during the phases of a strong dipole;
(3) corresponding hemispheric differences in the strength of proto-neutron 
star convection and in the electron and lepton fractions in the convection
layer and above it; (4) a hemispheric asymmetry of the accretion layer between
deformed shock and neutron star with greater accretion on the side of the
higher lepton-number flux; (5) SASI-induced modulations of the LESA
dipole in periods of strong SASI activity, when the LESA dipole is located
close to the direction of SASI sloshing or close to the plane of SASI 
spiral motions.

Our results clearly prove that LESA is {\em not} an artifact of the RbR+ 
approximation. Nevertheless, there are interesting differences between the
FMD and RbR+ results in certain aspects. The LESA dipole amplitudes with 
RbR+ neutrino transport tend to grow noticeably faster, reach higher 
amplitudes earlier, and their maximum amplitudes can, at least temporarily, 
be higher than those obtained with FMD transport. However, the dipole,
once fully developed, dominates the multipole spectra of the 
lepton-number flux in the FMD cases
in a much cleaner way than in the RbR+ runs (although it can be 
somewhat lower compared to the monopole). The spectrograms of the RbR+
models typically show prominent quadrupoles as well, and significant power
over a wider range of higher-order multipoles. This patchiness of the
directional variations of the lepton-number emission is expected as
a natural consequence of the more localized neutrino effects in the
RbR+ approximation (see Paper~I).
In contrast, the neutrino field leaving the convective layer is \new{smoothed} in the surrounding radiative shell
by non-local radiation-transport effects of FMD.

Closely inspecting the mass flows in and around the 
convection layer interior to the proto-neutron star during phases of
a strong LESA dipole, we reveal a global, volume-filling
dipolar pattern that streams
around the central, convectively stable core. These currents effectively
transport lepton-rich matter from the hemisphere opposite to the
LESA dipole direction to the hemisphere in the LESA dipole direction,
and from deeper regions to layers closer to the neutrinospheres.
An important effect is the gain or loss of electron lepton-number
by these currents through exchange with diffusion fluxes (via neutrino 
absorption or emission) when the gas flow moves along the inner and 
outer boundaries between the convective layer and the diffusive 
central core on the one side and the enveloping shell on the other
side. This exchange process fuels and
defuels the lepton content of the currents.

The presence of a dipolar flow pattern through and around the convection
zone of the neutron star motivated us to investigate whether this
finding is compatible with predictions by Chandrasekhar's linear
stability analysis of thermal instability in spherical shells
\citep{1961hhs..book.....C}.
Our assessment can, of course, only be tentative because of the
substantial differences between the complex physical conditions
in the Ledoux-unstable neutron-star layer compared to the idealized
terrestrial setups investigated by \citet{1961hhs..book.....C}.
Despite this caveat the growth behavior of the different multipoles
of the lepton-number flux, which we relate to multipoles of the
flow in the convective layer, follows amazingly well the 
expectations from Chandrasekhar's theory. We conclude this from
comparing the Rayleigh numbers of the system, which we estimate
at the center of the convective layer, at the times when we identify
the initial rise of an $l$-mode emission asymmetry, with the critical
Rayleigh number for this $l$-mode derived by Chandrasekhar.
His theory predicts that an $l$-mode perturbation should reach
marginal stability when the Rayleigh number grows to the critical
value of the mode. In agreement with the mathematical analysis we
witness the onset of modes around $l = 4$ first, and the growth
of the other multipole modes setting in at times when the Rayleigh
numbers ${\cal R}$ for the conditions of the system follow the 
critical curve ${\cal R}_{\mathrm{C},l}$ adopted from
\citet{1961hhs..book.....C}. Since the Rayleigh number of the
convective layer continues to rise monotonically also after
the dipole mode, $l=1$, appears for the first time, one can expect
its presence also at later times. 

Calculating the Rayleigh number requires the specification of
the kinematic viscosity of the environment. In current models 
viscous effects are
far dominated (by a factor $\sim$10--100) by numerical viscosity.
This needs to be kept in mind when numerical simulations are
evaluated for the growth of LESA multipole modes.

Our reported work has attempted to achieve progress towards a better
understanding of the puzzling LESA phenomenon. Despite the assuring
result that the existence of a dipolar flow pattern in the
neutron-star convection layer seems compatible with Chandrasekhar's
theory, the detailed reason for the dominance of the $l=1$ mode
remains unclear. Previous speculations that it is driven by feedback
effects with a hemispheric accretion asymmetry around the neutron
star \citep{2014ApJ...792...96T}, or by an amplifying feedback cycle
connected to $Y_\mathrm{lep}$-dependent differences of Ledoux
convection in opposite neutron star hemispheres 
\citep{2016ARNPS..66..341J}, require corroboration by future
investigations.

\new{Moreover, plenty of future work will have to address the consequences of LESA
in the context of supernova explosions and their detectable neutrino emission.
Interpreting our present results (and those of previous reports of LESA in the literature)
we would not think that LESA has any significant impact on the supernova mechanism
and the ``explodability'' of progenitor stars.
On the one hand, the LESA asymmetry concerns mainly the lepton-number flux
but much less the total (number and energy) flux of electron neutrinos and antineutrinos
(see \citealt{2014ApJ...792...96T,2014PhRvD..90d5032T}),
for which reason the emission asymmetry has a subordinated consequence on the neutrino heating in the post-shock layer.
LESA effects in this respect seem to be overruled by SASI- and convection-induced perturbations and modulations,
which is reflected by the fact that SASI activity in our 20\,$M_\odot$ runs has considerable implications
for the stability of the LESA direction and amplitude (see also \citealt{2014PhRvD..90d5032T}).
The assessment of probable implications of LESA for neutron star kicks and supernova nucleosynthesis
(as mentioned already by \citealt{2014ApJ...792...96T})
and possible implications for explosion energies will require much longer simulations than performed in our work.
While consequences of LESA for the detectable neutrino signal have been addressed by
\citet{2014PhRvD..90d5032T} as well as \citet{2019arXiv190106235W},
a potential connection of LESA and the gravitational-wave emission of newly formed proto-neutron stars
will have to be investigated in the future (but for a first glimpse, see \citealt{2018arXiv181205738P}).}

\acknowledgments

\new{We are grateful to Irene Tamborra for her valuable comments on the manuscript and to Bernhard M{\"u}ller for enlightening discussions.
We further thank Shaoming Zhang for his help with the 3D visualizations.}
At Garching, this project was supported
by the European Research Council through grant ERC-AdG No.\ 341157-COCO2CASA,
and by the Deutsche Forschungsgemeinschaft
through Sonderforschungbereich SFB 1258 ``Neutrinos and Dark Matter in Astro- and Particle Physics'' (NDM)
and the Excellence Cluster Universe (EXC 153; http://www.universe-cluster.de/).
OJ acknowledges support by the Special Postdoctoral Researchers (SPDR) program and the iTHEMS cluster at RIKEN.
Computer resources for this project have been provided by the Leibniz Supercomputing Centre (LRZ) under grant pr62za,
and by the Max Planck Computing and Data Facility (MPCDF) on the HPC system Hydra.

\software{\textsc{Aenus-Alcar} \citep{{2008ObergaulingerPhD},2015MNRAS.453.3386J,2018MNRAS.481.4786J},
NumPy and Scipy \citep{Oliphant2007},
IPython \citep{Perez2007},
Matplotlib \citep{Hunter2007},
\new{VisIt \citep{Childs2012}}.}

\bibliographystyle{abbrvnat}

\end{document}